\documentclass[a4paper, onecolumn]{article}

\usepackage{geometry}
\usepackage[english]{babel}
\usepackage[utf8]{inputenc}
\usepackage{booktabs}
\usepackage[T1]{fontenc}
\usepackage{array}

\usepackage{authblk}
\usepackage{graphicx}
\usepackage[labelfont=bf, figurename=Fig., font={small}]{caption}
\usepackage{natbib}

\usepackage[group-four-digits=true,
            output-decimal-marker={.}, 
            ]{siunitx}
\usepackage{listings}

\usepackage{makecell} 

\usepackage{float}
\usepackage{fancyhdr}
\usepackage{xr-hyper}
\usepackage{hyperref}
\hypersetup{
    colorlinks=True,
    linkcolor=blue,
    filecolor=black,     
    urlcolor=blue,
    anchorcolor=black,
    citecolor=black
    }

\usepackage[noabbrev,capitalize]{cleveref}

\usepackage{setspace}

\usepackage[group-four-digits=true,
            output-decimal-marker={.},
            ]{siunitx}
\usepackage{float}
            
\usepackage{caption}
\usepackage{subcaption}

\title{Network analysis of the Danish bicycle infrastructure: Bikeability across urban-rural divides}

\author[1]{Ane Rahbek Vierø}
\author[1,2]{Michael Szell}
\affil[1]{Networks, Data, and Society (NERDS), Computer Science Department, IT University of Copenhagen, 2300 Copenhagen, Denmark} 
\affil[2]{Complexity Science Hub Vienna, 1080 Vienna, Austria}

\date{}

\begin{document}

\maketitle

\begin{abstract}
\noindent Research on cycling conditions focuses on cities, because cycling is commonly considered an urban phenomenon. People outside of cities should, however, also have access to the benefits of active mobility. To bridge the gap between urban and rural cycling research, we analyze the bicycle network of Denmark, covering around \num[round-mode=places,round-precision=0]{43000} \si{km^{2}} and nearly 6 mio.~inhabitants. We divide the network into four levels of traffic stress and quantify the spatial patterns of bikeability based on network density, fragmentation, and reach. We find that the country has a high share of low-stress infrastructure, but with a very uneven distribution. The widespread fragmentation of low-stress infrastructure results in low mobility for cyclists who do not tolerate high traffic stress. Finally, we partition the network into bikeability clusters and conclude that both high and low bikeability are strongly spatially clustered. Our research confirms that in Denmark, bikeability tends to be high in urban areas. The latent potential for cycling in rural areas is mostly unmet, although some rural areas benefit from previous infrastructure investments. To mitigate the lack of low-stress cycling infrastructure outside of urban centers, we suggest prioritizing investments in urban-rural cycling connections and encourage further research in improving rural cycling conditions.
\end{abstract}

\bigskip


\section{Introduction}

Despite cycling often being framed as an urban phenomenon \citep{kircher_cycling_2022}, it is crucial that cycling research and planning also address rural and suburban areas, as cycling requires safe and well connected bicycle infrastructure \emph{everywhere}. Many (potential) bicycle commuters travel across urban-suburban-rural divides \citep{skov-petersen_effects_2017, anderson_dataanalyse_2019} and the growing prevalence of e-bikes makes longer distance cycle trips feasible for a growing part of the population \citep{hallberg_modelling_2021}. Furthermore, cycling is a healthy, accessible, and inexpensive mode of transport that can counterbalance a lack of access to other transport modes \citep{mueller_health_2018, lee_understanding_2017}. Cycling should thus also be accessible to people living outside of cities. Cycling conditions nevertheless tend to be worse in rural areas, and cycling research on non-urban cycling conditions is scarce \citep{kircher_cycling_2022}. 

\begin{figure}[H]
\centering
\includegraphics[width=0.99\textwidth]{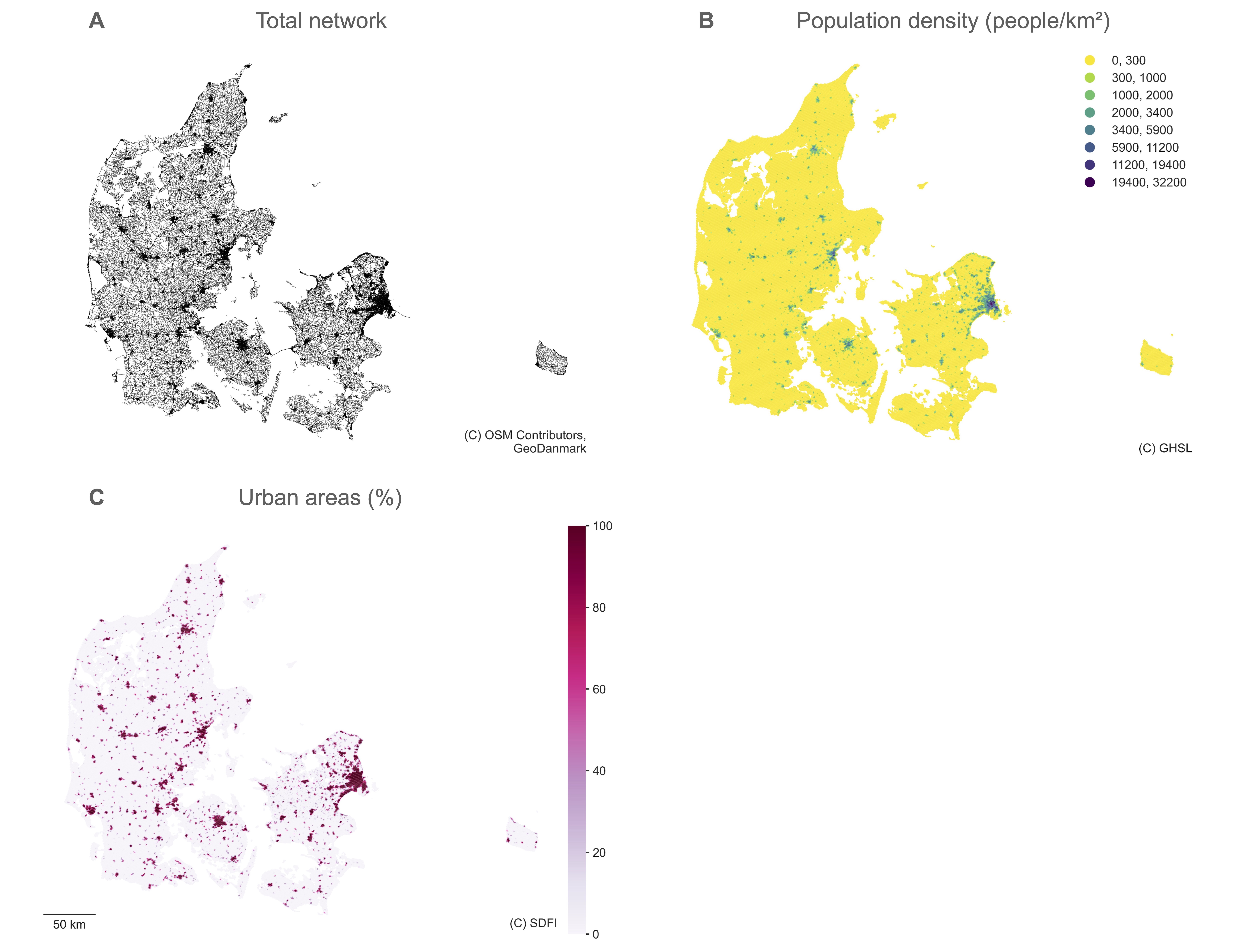}
\caption{\textbf{Study area: Denmark.} A) The total Danish road and path network. B) Population density. C) Urban areas aggregated at a hex grid level, as classified by the Danish Agency for Climate Data and Danmarks Miljøportal (The Danish Environmental Portal).}
\label{fig:overview}
\end{figure}

Going across urban-rural divides, in this study we analyze the cycling conditions, or `bikeability', for the entire country of Denmark, covering \num[round-mode=places,round-precision=0]{43057} \si{km^{2}} (including $\sim$\num[round-mode=places,round-precision=0]{3000} \si{km^{2}} urban areas), and nearly \num[round-mode=places,round-precision=0]{6} mio.~inhabitants (Fig.~\ref{fig:overview}). Denmark is known for its comparatively high cycling rates and strong cycling culture \citep{christiansen_international_2016, carstensen_cycling_2012, rich_our_2023}. However, cycling rates for children and people living outside of the largest cities have been declining, while general cycling levels are stagnating \citep{rich_our_2023, christiansen_danish_2023} and car ownership is on the rise \citep{statistics_denmark_mere_2023}.

Recent research suggests that there is a great potential for converting shorter commuter trips from cars to bicycles \citep{schmidt_identifying_2024}. Converting cycling potential to actual cyclists is a complex issue \citep{assuncao-denis_increasing_2019, xiao_shifting_2022} but one important factor is living in an area with adequately safe and comfortable bicycle infrastructure, where bicycle infrastructure denotes the road and path network open to cyclists \citep{buehler_bikeway_2016, cervero_network_2019, kamel_impact_2021, fosgerau_bikeability_2023}.

To understand \textit{where} and to \textit{what extent} the current bicycle infrastructure is conducive to cycling, we use OpenStreetMap (OSM) data, enriched with the national public data set GeoDanmark, to evaluate the local network density, network fragmentation, and network reach for different configurations of the bikeable road network based on the concept of `Levels of Traffic Stress' \citep{mekuria_low-stress_2012}. We examine the results of the network analysis for spatial patterns and identify clusters of low and high bikeability, to answer the research question: \textit{What are the spatial patterns in bikeability across Denmark?}

We find that, at the country level, Denmark has a relatively high \textit{share} of low-stress infrastructure, but it is very unevenly distributed with most low-stress infrastructure clustered in a few locations. This heterogeneity implies that a large part of the country's population does not live in close proximity to the low-stress infrastructure necessary to support higher cycling rates. The networks of lower stress infrastructure, in addition, display a much higher network fragmentation than networks including higher stress infrastructure, resulting in limited mobility for cyclists unwilling or unable to bike on high-stress roads. While a lack of safe and low-stress bicycle infrastructure affects all cyclists, it is known to affect some population groups in particular, such as children, the elderly, inexperienced cyclists, and people with disabilities \citep{clayton_cycling_2017, doran_pursuit_2021}. Although we find a tendency for areas with high population densities to have a higher bikeability, we also identify rural areas with a high share of low-stress infrastructure or low-stress connections over longer distances outside of urban areas. This finding illustrates that good cycling conditions do not have to be restricted to high-density urban areas.

\bigskip 

The rest of the paper is organized as follows: first, we provide an introduction to previous work on levels of traffic stress, bikeability, and cycling in non-urban areas (Section \ref{section:review}) and provide an overview of the input data and methods in this study (Section \ref{section:data_methods}). We then present the results of the analysis (Section \ref{section:results}), followed by a discussion of the findings and limitations of the study (Section \ref{section:discussion}), and finally we summarize the main findings on the bikeability of the Danish road network in the conclusion (Section \ref{section:conclusion}).
 
\section{Literature review}
\label{section:review}

The field of bicycle research has been in rapid development in the past decade, with an especially large increase in research projects that use quantitative methods to examine cycling conditions \citep{buehler_bikeway_2016, castanon_bikeability_2021}. In the literature review below, we limit ourselves to work focusing on bicycle networks, whether defined as networks of dedicated bicycle infrastructure (i.e.~bicycle tracks, lanes, etc.) or the entire road network. The review is divided into three sections, although the topics are highly interconnected: first, we introduce the idea of `bicycle suitability' and Levels of Traffic Stress (LTS); second, we provide an overview of `bikeability' and bicycle network analysis; and third, we discuss the state of research on cycling conditions outside of urban areas.

\subsection{Bicycle suitability \& LTS}
\label{subsection:suitability-lts}

The first step towards understanding how conducive a place is to cycling is to model bicycle suitability at the network link level. The concept of `bicycle suitability' was suggested by \cite{lowry_assessment_2012} to capture how well suited a specific road or path segment is for cycling. Bicycle suitability is closely related to concepts such as Bicycle Level of Service (BLOS) and Levels of Traffic Stress (LTS), which all classify each network segment based on attributes such as the presence of dedicated bicycle infrastructure, traffic volumes or speed limits ~\citep{lowry_assessment_2012, mekuria_low-stress_2012, arellana_developing_2020}. Particularly the concept of LTS has gained popularity in research \citep{wang_investigating_2022}. The LTS framework classifies the road and path network available to cyclists into four classes based on characteristics such as the presence of protected bicycle infrastructure, posted speed limits, traffic volumes, and on-street parking. LTS 1 is assumed safe and low-stress enough for all cyclists, including children, while LTS 4 only is suitable for the "strong and fearless" \citep{mekuria_low-stress_2012, furth_network_2016}. 

The LTS classification has received some criticism for a lack of empirical foundation and mixed results for predicting cycling levels \citep{buehler_bikeway_2016, wang_does_2016, bearn_adaption_2018, cabral_empirical_2022}. Bicycle networks with low LTS are however associated with higher cycling levels and streets with a low LTS score have less severe bicycle crashes \citep{chen_how_2017, cervero_network_2019}. The core assumptions in the LTS classification -- that cyclists prefer protected bicycle infrastructure and roads with low traffic volumes and low speed limits -- moreover aligns with most findings on cycling route choice and stated preferences \citep{winters_motivators_2011, dill_revisiting_2016, buehler_bikeway_2016, vedel_bicyclists_2017, gossling_subjectively_2022, lukawska_joint_2023}. The LTS classification scheme is thus generally accepted as a useful method for understanding bicycle suitability at the link level \citep{wasserman_evaluating_2019, wang_investigating_2022}.

Although LTS originally was associated with a predefined set of criteria \citep{sorton_bicycle_1994, mekuria_low-stress_2012}, numerous efforts have been made to adapt and simplify the classification method \citep{semler_low-stress_2017, wasserman_evaluating_2019, cabral_empirical_2022,wang_investigating_2022}. The original LTS classification relies on a large number of variables that are often not available to neither bicycle planners nor researchers \citep{bearn_adaption_2018, wang_investigating_2022}. To make the LTS framework applicable in locations without the data required by the original classification method, recent research has aimed at simplifying the criteria and examining the accuracy of LTS scores computed with fewer data points \citep{semler_low-stress_2017, bearn_adaption_2018, wasserman_evaluating_2019, crist_fear_2019, wang_investigating_2022}. Many variables, such as number of lanes and posted speed limits, are highly correlated and can often be imputed based on road type, and thus allow successful LTS classifications despite missing data \citep{semler_low-stress_2017, bearn_adaption_2018, wasserman_evaluating_2019, wang_investigating_2022}. 

Of particular interest to this study are the findings from \cite{wasserman_evaluating_2019} who evaluate the feasibility of using OSM data to predict LTS in Montgomery, MA. They conclude that a combination of OSM tags and imputation of missing data based on road type correctly can predict low or high LTS ratings for almost \num[round-mode=places,round-precision=0]{90}\% of the network, but also that the prediction accuracy varies with road type and population density, with higher accuracy in high-density areas. \cite{wang_investigating_2022} similarly find that the road type and derived attributes are well suited to predict LTS, but similarly conclude that LTS scores based on partial data sets were less accurate in rural areas. Both \cite{wasserman_evaluating_2019} and \cite{wang_investigating_2022} emphasize the importance of adapting LTS classifications and missing data procedures to the local context, just as \cite{bearn_adaption_2018} point out that the original LTS classification is based on Dutch guidelines and thus may need to be adapted when used in locations with different cycling cultures. In further development of the core idea behind LTS and other bicycle suitability classifications, \cite{reggiani_multi-city_2023} classify the road network based on the presence and type of bicycle infrastructure and road type, while \cite{winters_at--glance_2020, winters_canadian_2022} and \cite{ferster_developing_2023} stratify the Canadian network of dedicated bicycle infrastructure based on bicycle comfort and safety.

\subsection{Bikeability \& bicycle networks}

While LTS rankings can tell us something about bicycle suitability at the link level, `bikeability' is used to denote bicycle friendliness at an aggregated scale. The concept of bikeability is widely used, however, with no universally accepted definition \citep{arellana_developing_2020, kellstedt_scoping_2021}. Bikeability addresses how suitable an area is for cyclists based on both measures of \textit{comfort} and \textit{safety} \citep{reggiani_understanding_2022}, but with large variations in the spatial scale of measurement \citep{nielsen_bikeability_2018, arellana_developing_2020, werner_bikeability_2024}. \cite{lowry_assessment_2012} suggest measuring bikeability at the \textit{network level} based on an assessment of cyclists' comfort and convenience when accessing important destinations. The network focus in bikeability is widely used and has been extended by e.g.~\cite{winters_mapping_2013}, who also include network connectivity, topography, and land use, and \cite{nielsen_bikeability_2018} who additionally incorporate regional variables believed to influence the local bicycle mode share. More recently, \cite{reggiani_understanding_2022} has assessed urban bikeability based on the comfort and directness of bicycle connections at the city zone level, while \cite{fosgerau_bikeability_2023} use the concept in an assessment of cyclist preferences and the effect of bicycle infrastructure on cycling levels. 

Although the many different ways of measuring bikeability make it unfeasible to summarize any clear findings, previous studies have concluded that bikeability levels vary substantially between different locations, even within the same city, and that areas with high bikeability ratings tend to have higher levels of cycling \citep{winters_mapping_2013, nielsen_bikeability_2018, reggiani_understanding_2022, wysling_where_2022, fosgerau_bikeability_2023, beecham_connected_2023}. In this study we measure bikeability based on key indicators for network density and connectivity for different levels of LTS (Section \ref{section:data_methods}). To go from LTS and bicycle suitability to bikeability, we thus need to move the focus from the network link to the network structure.

Within research on the structural properties of bicycle networks, one line of studies focuses on assessing the correlation between cycling levels and bicycle network quality \citep{schoner_missing_2014, dill_bicycle_2003, buehler_cycling_2012, kamel_impact_2021}, while others, as in this study, limit their focus to evaluations of bicycle networks themselves \citep{vybornova_automated_2022, reggiani_multi-city_2023}. Metrics used to evaluate bicycle networks include, for example, network and intersection density, link/node ratio,  directness/circuity, network centrality, component size, network gaps, and missing links \citep{dill_measuring_2004, kamel_impact_2021, reggiani_multi-city_2023, schon_scoping_2024}.

There is a general consensus that not only the length but also the structure and connectivity of local networks of dedicated bicycle infrastructure matters for cycling levels \citep{dill_bicycle_2003, buehler_cycling_2012,  schoner_missing_2014, buehler_bikeway_2016, fosgerau_bikeability_2023, kamel_impact_2021, schon_impact_2024}. Similarly, stated-preference research shows that most cyclists prefer separated, continuous, and connected bicycle infrastructure \citep{buehler_bikeway_2016}. Conversely, it has been shown that poor network quality such as sudden network discontinuities increase cyclists' discomfort \citep{krizek_what_2005}. Although exact findings on bicycle network quality vary from location to location and depend on the definition of `bicycle network' \citep{reggiani_multi-city_2023, schon_scoping_2024}, several studies have identified how networks of dedicated bicycle infrastructure tend to suffer from discontinuities and missing links \citep{krizek_what_2005, vassi_review_2014,  vybornova_automated_2022}. Bicycle networks are usually much more fragmented than the regular street network with many disconnected components and isolated islands of low-stress streets \citep{schoner_missing_2014, natera_orozco_data-driven_2020, reggiani_multi-city_2023}. This fragmentation either prevents people from cycling or force cyclists onto mixed traffic and high-stress streets \citep{furth_network_2016, semler_keys_2018, crist_fear_2019}.

\subsection{Cycling in non-urban areas}

Although bicycle networks are increasingly well studied, there is a large knowledge gap when it comes to cycling conditions outside of urban areas, as cycling is often framed as an urban phenomenon \citep{aytur_pedestrian_2011, mcandrews_reach_2017, kircher_cycling_2022}. That cycling is associated with urban settings is partly justified: we know that urban environments with dense street networks and proximity to many destinations are conducive for active mobility \citep{dill_bicycle_2003, berrigan_associations_2010, kamel_impact_2021, nielsen_bikeability_2018}. Furthermore, delimited urban areas with higher population densities are often more suitable for research projects that aim to correlate bicycle conditions with cycling levels. However, this focus leaves a large gap both in terms of \textit{knowledge} of the state of non-urban cycling networks and \textit{methods} for studying non-urban cycling \citep{mcandrews_motivations_2018, kircher_cycling_2022, scappini_regional_2022}. For example, many traditional network metrics are not suitable for disconnected networks, or are highly correlated with urban density and the size of the study area \citep{borgatti_centrality_2005, knight_metrics_2015, marshall_street_2018}, and thus cannot be readily applied on larger study areas with much sparser networks.

Previous work on cycling conditions outside of urban areas focuses mainly on adapting bikeability indexes to rural contexts \citep{noel_compatibility_2003, jones_development_2003}, examining rural bicycle planning practices \citep{mcandrews_motivations_2018}, planning for recreational cycling \citep{scappini_regional_2022}, or is published as reports with a general focus on promoting cycling outside of urban areas \citep{gardner_preliminary_1998, kircher_cycling_2022}. While cycling indeed is more prevalent in dense urban environments \citep{mcandrews_reach_2017, nielsen_bikeability_2018}, there is substantial cycling potential in rural areas \citep{kircher_cycling_2022, schmidt_identifying_2024}. Despite indications that cycling in rural areas can be more dangerous \citep{macpherson_urbanrural_2004}, cycling rates per population are moreover comparable between urban and rural settings in some locations \citep{tribby_examining_2019}.
In Denmark, people outside of urban areas are however less likely to commute by bicycle compared to urban residents, even for those living less than 5 km from their workplace \citep{schmidt_identifying_2024}. 

The low rates of active mobility outside urban areas have been problematized from a climate and sustainability perspective \citep{peer_which_2023, leichenko_promoting_2024}. However, the tendency to focus cycling research and planning efforts on urban areas is not just a climate concern, but also results in spatial inequities in access to the bicycle as a cheap and healthy mobility mode. The increasing prevalence of electric bicycles and cycle highways (longer-distance cycle paths designed especially for fast and low-stress commuter cycling) could help change current mobility patterns by enabling longer and faster bicycle trips \citep{bourne_impact_2020, hallberg_modelling_2021}. At present, however, there is a missed potential for more active mobility in urban and non-urban areas alike.

\section{Data \& methods}
\label{section:data_methods}

In the following section, we first introduce the data used in the study, then introduce the classification of the network into different levels of bicycle suitability, and end with an overview of the applied bikeability metrics.

\subsection{Input data}

The study makes use of four input data sets:

\begin{itemize}
    \item OSM road network data.
    \item GeoDanmark bicycle network data.
    \item Population density data.
    \item Area data for urban zones across Denmark.
\end{itemize}

The main data set is data on the entire Danish road network from the global, crowd-sourced mapping platform OpenStreetMap (OSM), downloaded from Geofabrik in July \num[round-mode=places,round-precision=0, group-separator = {}]{2024} \citep{geofabrik_our_2020}. To compensate for issues with incomplete OSM data, particularly data on dedicated bicycle infrastructure \citep{hochmair_assessing_2015, ferster_using_2020, viero_how_2024}, the OSM data are enriched with data on bicycle tracks and lanes from the national public data set GeoDanmark \citep{geodanmark_danmarks_2023} using the methods developed by \cite{viero_bikedna_2024}. The study furthermore includes data on population densities from a high-resolution population raster from the European Union's Global Human Settlement Layer \citep{schiavina_ghs-pop_2023} and data on urban zones \citep{sdfi_danske_2024, danmarks_miljoportal_planlaegning_2024}.

\subsection{Classification of bicycle suitability}
\label{subsection:lts-classi}

An unresolved issue for research on `bicycle networks' is how much of the road and path network a bicycle network does or should include \citep{van_der_meer_assessment_2024, schon_scoping_2024}. Some studies include only dedicated bicycle facilities such as bicycle lanes or tracks \citep{dill_bicycle_2003, buehler_cycling_2012, schoner_missing_2014, kamel_impact_2021, vybornova_automated_2022, szell_growing_2022}, some include all parts of the road network considered low-stress \citep{furth_network_2016, lowry_quantifying_2017, semler_keys_2018}, while others include all parts where cycling is allowed \citep{dill_measuring_2004}, or combine different definitions of bicycle networks \citep{reggiani_multi-city_2023, van_der_meer_assessment_2024}. The appropriate delineation will always be context-specific \citep{schoner_missing_2014} and depend on e.g.~local regulations, cycling culture, and traffic safety. Limiting the bicycle network to designated bicycle infrastructure might underestimate the actual bicycle conditions where cycling in mixed traffic is appropriate \citep{schoner_missing_2014}. Conversely, including all roads where cycling is legal will fail to capture the constraints experienced by cyclists who are not comfortable cycling in mixed traffic, especially in areas with high traffic volumes or high speeds. A popular solution is to stratify the road network into different levels of assumed bicycle suitability, either based on road type \citep{reggiani_multi-city_2023}, bicycle facility type \citep{winters_at--glance_2020}, or more complex classifications such as LTS (Section \ref{subsection:suitability-lts}). In this study, we apply the later approach by including the entire network where cycling is allowed, but classified into four levels of LTS.

\begin{figure}[H]
\centering
\includegraphics[width=0.99\textwidth]{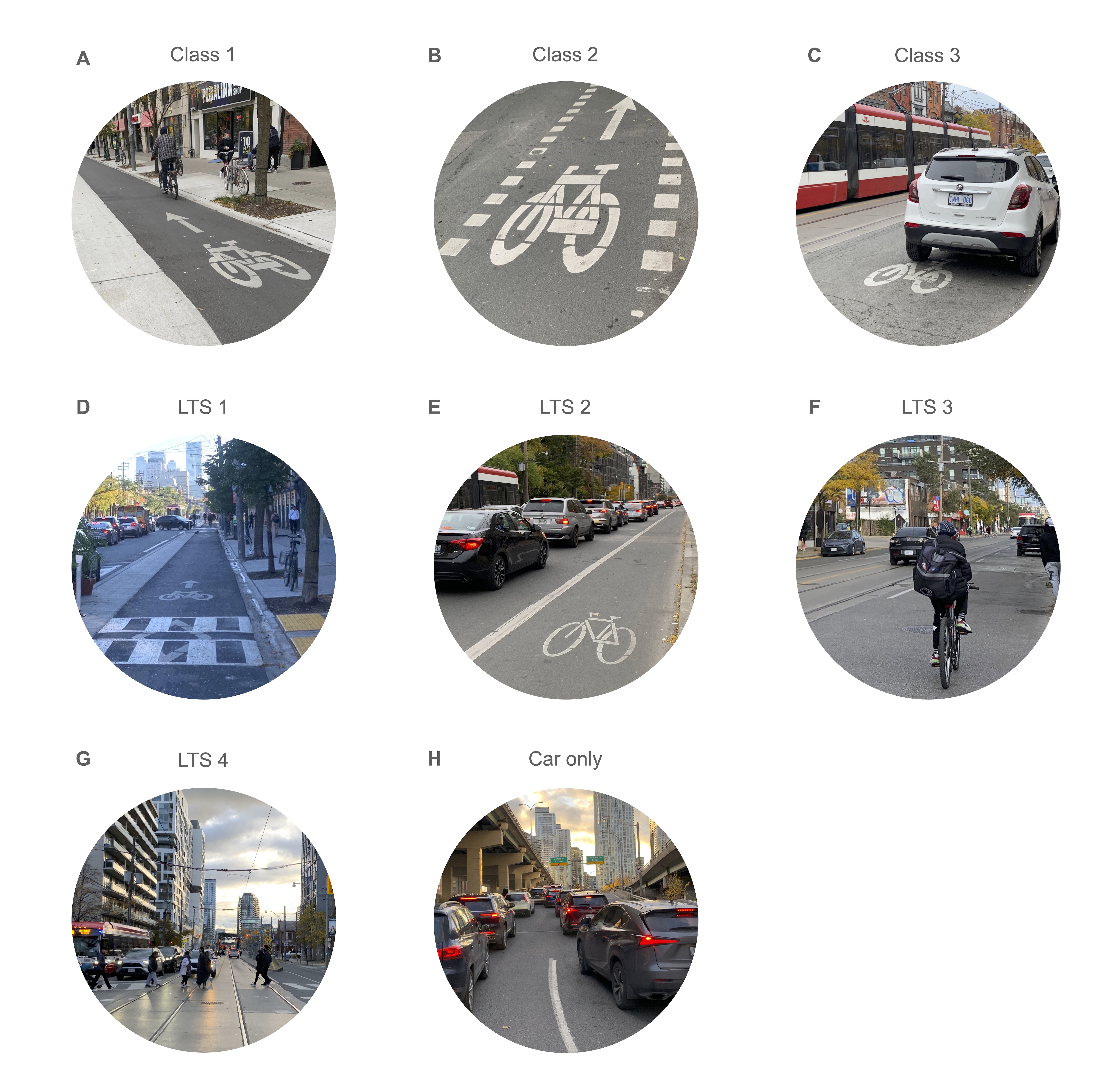}
\caption{\textbf{Examples of bicycle classes and LTS classifications.} A) Bicycle class 1: Protected, separated bicycle infrastructure. B) Bicycle class 2: Unprotected bicycle infrastructure. C) Bicycle class 3: Cyclists are mixed with motorized traffic. D) LTS 1. E) LTS 2. F) LTS 3. G) LTS 4. H) Motorway and other car-only roads are not included in the bikeable network.}
\label{fig:bicycle-classes-lts-examples}
\end{figure}

We classify the road network into four different levels, in accordance with the traditional LTS classification \citep{mekuria_low-stress_2012, furth_network_2016}, where 1 represents the lowest level of traffic stress (highest level of cycling suitability) and 4 the highest level of traffic stress (lowest level of cycling suitability) (Table \ref{table:lts-classi} and Fig.~\ref{fig:bicycle-classes-lts-examples}). LTS 1 and 2 are considered low-stress, while LTS 3 and 4 are considered high-stress. We also include the part of the road network that is exclusively for car traffic to allow a comparison of bicycle and car connectivity. Car-only roads are highways and similar roads where cycling is not allowed, as well as any road with a separate bicycle track in which case the bicycle track must be used. Paths with surfaces unsuitable for everyday cycling (grass, sand, clay, etc.) are not included in the analysis.

Classifying a road network into LTS is a two-step process. In the first step, the bicycle network is divided into three different bicycle classes \citep{wasserman_evaluating_2019} (Table \ref{table:lts-classi}, Fig.~\ref{fig:bicycle-classes-lts-examples}A-C):

\begin{itemize}
    \item \textit{Bicycle class 1}: Protected bicycle tracks and bicycle paths separated from motorized traffic.
    \item \textit{Bicycle class 2}: Bicycle lanes and shared bicycle and bus lanes without physical separation of cyclists and motorized traffic.
    \item \textit{Bicycle class 3}: Any road where cyclists are cycling in mixed traffic.
\end{itemize}

In the second step, each road segment is classified according to the LTS criteria (Table \ref{table:lts-classi}, Fig.~\ref{fig:bicycle-classes-lts-examples}D-G). The variables in the LTS classification are:

\begin{itemize}
    \item \textit{Bicycle class (1, 2, or 3).}
    \item \textit{Posted speed limit.}
    \item \textit{Road type.}
    \item \textit{Number of road lanes.}
    \item \textit{Presence of bus route.}
\end{itemize}

\begin{table}[H]
\centering
\begin{tabular}{cp{3cm}p{3cm}p{3cm}p{3cm}}
\toprule
 \textbf{Bicycle class} & \textbf{LTS 1} & \textbf{LTS 2} & \textbf{LTS 3} & \textbf{LTS 4} \\
  \midrule
  Class 1   & \multicolumn{3}{c}{\textit{Class 1 is always LTS 1}}   \\
\midrule
  Class 2  & speed limit $\leq$ 50 AND lanes $\leq$ 2 & Speed limit $\leq$  50 AND lanes $\leq$ 4 OR speed limit $<$  50 AND bus route & speed limit $\leq$ 60 AND lanes $\leq$ 4 OR speed limit $<$ 70 AND bus route & speed limit $\geq$ 70 OR lanes $\geq$ 5 \\
\midrule  
  Class 3 &  speed limit $\leq$ 30 AND lanes $\leq$ 2 OR speed limit $\leq$ 20 h AND lanes $\leq$ 3 & Speed limit $\leq$ 50 AND lanes  $\leq$ 3 OR speed limit $\leq$ 30 AND bus route & speed limit $\leq$ 50 AND lanes $\leq$ 4 OR speed limit $\leq$ 50 AND bus route OR speed limit $\leq$ 50 AND road type IN (‘primary’, ‘secondary’, ‘tertiary’) & Speed limit $\geq$ 50 OR lanes $\geq$ 4 OR speed limit $>$ 50 AND bus route \\
\bottomrule
\end{tabular}
\smallskip
\caption{\textbf{LTS classification criteria}. All criteria for a given level must be fulfilled to achieve that level of traffic stress. If not, the segment is moved to a higher LTS class. Speed limits are in km/h. Roads and paths without public access are not included.}
\label{table:lts-classi}
\end{table}

This LTS classification is inspired by the OSM compatible criteria developed by \cite{wasserman_evaluating_2019}, but further simplified and adjusted to a Danish context and local data availability, in line with recommendations from previous research on LTS adaptations \citep{bearn_adaption_2018, wasserman_evaluating_2019, wang_investigating_2022}. Local adaptations, for example, concern speed limits (adapted to match Danish speed limits) and road center lines (not included due to insufficient data). Road type and posted speed limits are used as a proxy for traffic volumes, in line with related research using OSM data for LTS analysis \citep{wasserman_evaluating_2019, crist_fear_2019, wang_investigating_2022, werner_bikeability_2024}. 

Although OSM road network data are generally highly annotated with a large number of tags (attributes), OSM also suffers from incomplete data and missing tags, especially in areas with lower population densities \citep{barrington-leigh_worlds_2017, fonte_assessing_2017,viero_how_2024}. In this study, we interpolate missing data for speed limits and number of lanes based on the road type and whether the road is in an urban area or not (roads are classified as urban or non-urban based on the intersection with urban zone data). Although this type of data interpolation naturally introduces some uncertainty in the results, recent research on data interpolation in LTS classifications has demonstrated that missing data on number of lanes, speed limits, and traffic volumes can be inferred from road type with high accuracy \citep{wang_investigating_2022}.

To ensure the most accurate classification possible, the classified road network data set was shared publicly on an online map from February to June 2024 with the possibility of adding comments and feedback on the input data and the assigned LTS class \citep{viero_bicycle_2024}, resulting in 35 responses from cyclists, municipal bicycle planners, and bicycle NGOs. The feedback process, for example, identified locations missing a bicycle track in OSM and resulted in the exclusion of some roads and paths with surfaces unsuitable for regular cyclists. The feedback also resulted in including bus routes as a criterion in the LTS classification, as many responses identified roads with regular bus routes as high-stress despite relatively low speed limits. Finally, the classification of the Danish OSM data required extensive manual edits of the OSM database, particularly to identify roads with separately mapped bicycle tracks, in which case the main road geometry should not be included in the bikeable network (Fig.~\ref{SI-fig:osm-update-example}).

\begin{figure}[H]
\centering
\includegraphics[width=0.99\textwidth]{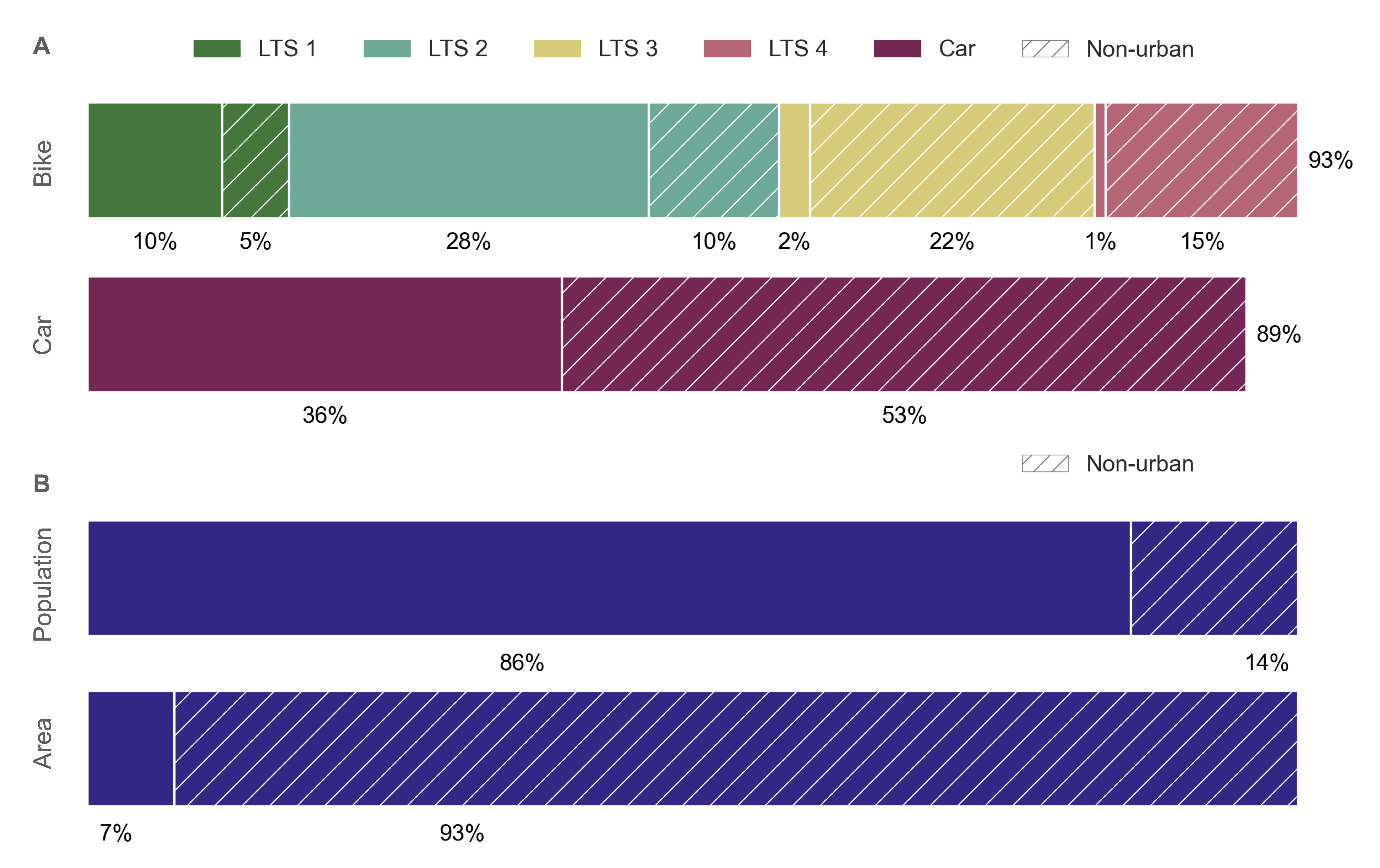}
\caption{\textbf{Overview of LTS shares, population and area distribution.} A) Share of network length in each LTS category in urban and non-urban areas. The bikeable network and the car network mostly overlap, but paths and separate bicycle tracks are not included in the car network, while highways and other non-cycling roads are excluded from the bikeable network. B) Share of population and area between urban and non-urban areas.}
\label{fig:lts-share}
\end{figure}

The entire Danish  network is \num[round-mode=places,round-precision=0]{130213.79} km long (Table \ref{table:lts-overview}). The total network includes all roads with public access, all bicycle tracks and lanes, and paths where cycling is not forbidden and where the path surface is conducive to utilitarian cycling (see Section \ref{SI-section:network-definition} in the Supplementary Information for further details). Applying the LTS classification (Table \ref{table:lts-classi}) to the network returns \num[round-mode=places,round-precision=0]{15.49}\% as \mbox{LTS 1}, \num[round-mode=places,round-precision=0]{37.65}\% as \mbox{LTS 2}, \num[round-mode=places,round-precision=0]{24.26}\% as \mbox{LTS 3},  and \num[round-mode=places,round-precision=0]{15.65}\% as \mbox{LTS 4} (Fig.~\ref{fig:lts-share}A and Table \ref{table:lts-overview}). LTS 1-4 make up the entire bikeable network. Of the included  network, \num[round-mode=places,round-precision=0]{93.05}\% allows cycling, and \num[round-mode=places,round-precision=0]{89.05}\% is open to cars. The non-bikeable part of the network consists of highways and other car-only roads, as well as roads with separate bicycle infrastructure (in which case cyclists must use the bicycle infrastructure, not the main road). The non-car part of the network consists of bicycle tracks, paths, and roads closed to motorized traffic, such as pedestrian and living streets.

\begin{figure}[H]
\centering
\includegraphics[width=0.99\textwidth]{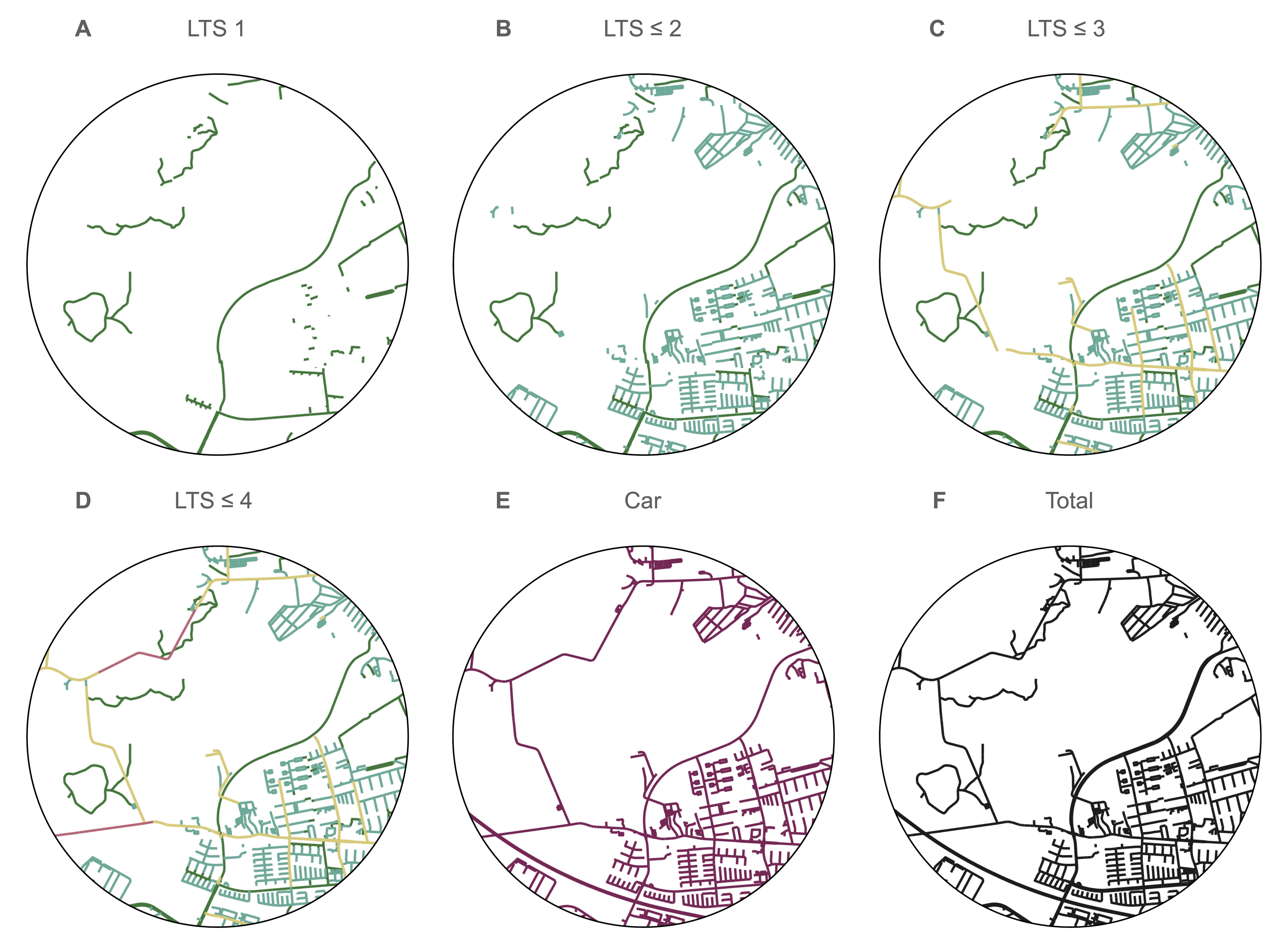}
\caption{\textbf{LTS networks.} A) LTS 1. B) LTS$\leq$2. C) LTS$\leq$3. D) LTS$\leq$4. E) Car. F). Full network. The LTS 1 network only include LTS 1 infrastructure, whereas LTS$\leq$2 include LTS 1-2, LTS$\leq$3 include LTS 1-3, and LTS$\leq$4 include LTS 1-4. The car network include all network links which allow for cars and have public access. The full network include the combination of the bikeable and the car network.}
\label{fig:lts-subgraphs}
\end{figure}

Each increasing LTS forms its own network from the total network. When aggregated from the bottom up, these sub-networks form a nested structure: We define each ``LTS$\leq$'' sub-network to include the network links from all lower levels of traffic stress, thereby extending them: the LTS 1 network only includes LTS 1 infrastructure, the LTS$\leq$2 network includes both LTS 1 and 2 infrastructure, the LTS$\leq$3 network includes LTS 1, 2, and 3 infrastructure, and the LTS$\leq$4 network includes LTS 1, 2, 3, and 4 infrastructure (Fig.~\ref{fig:lts-subgraphs}).

\begin{table}[H]
\centering
\begin{tabular}{lrr}
\toprule
\textbf{Network level}& \textbf{Length (km)} &\textbf{Share (\%)} \\
\midrule
LTS 1 & \num[round-mode=places,round-precision=0]{20163.91} & \num[round-mode=places,round-precision=0]{15.49} \\
LTS 2 & \num[round-mode=places,round-precision=0]{49028.83} & \num[round-mode=places,round-precision=0]{37.65} \\
LTS 3 & \num[round-mode=places,round-precision=0]{31592.05} & \num[round-mode=places,round-precision=0]{24.26} \\
LTS 4 & \num[round-mode=places,round-precision=0]{20377.99} & \num[round-mode=places,round-precision=0]{15.65} \\
Total bicycle & \num[round-mode=places,round-precision=0]{121162.78} & \num[round-mode=places,round-precision=0]{93.05} \\
Total car & \num[round-mode=places,round-precision=0]{115955.84} & \num[round-mode=places,round-precision=0]{89.05} \\
Total & \num[round-mode=places,round-precision=0]{130213.79} & \num[round-mode=places,round-precision=0]{100.00} \\
\bottomrule
\end{tabular}
\smallskip
\caption{\textbf{LTS network distribution}. Length and network share for each LTS level, the total bikeable network, total car network, and combined bikeable and car network.}
\label{table:lts-overview}
\end{table}

\subsection{Measuring bikeability}

The evaluation of bikeability, explained below, is based on three different network metrics:

\begin{itemize}
    \item \textit{Density.}
    \item \textit{Fragmentation.}
    \item \textit{Reach.}
\end{itemize}

All metrics are computed for each LTS to allow for comparison of bicycle network quality at different tolerances for traffic stress. Network density is computed separately for each separate LTS, while fragmentation and reach are computed for the LTS 1, LTS$\leq$2, LTS$\leq$3, LTS$\leq$4, and car networks, respectively. All results are aggregated at the local scale using a H3 hex grid \citep{uber_h3-py_2023} at resolution 8, with an average grid size of 0.7 \si{km^{2}} and \num[round-mode=places,round-precision=0]{78169} hex cells for the study area. Each hex cell is assigned an urban percentage based on the share of the hex cell that intersects an urban zone.

\begin{figure}[H]
\centering
\includegraphics[width=0.99\textwidth]{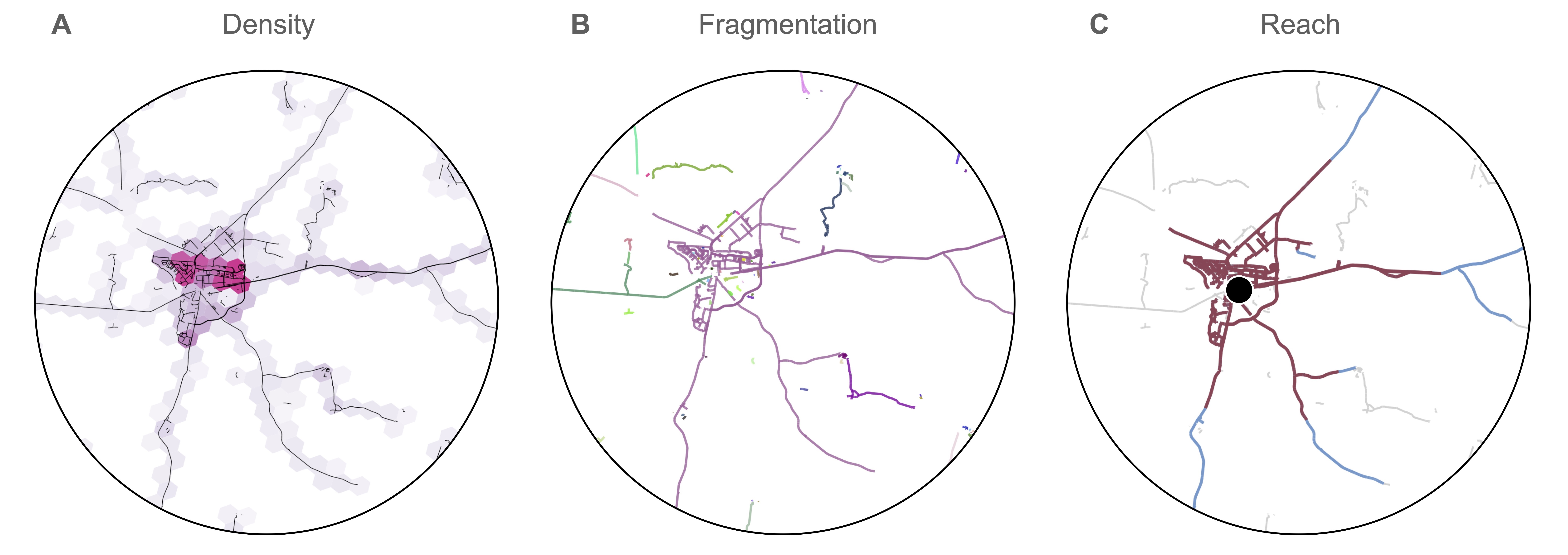}
\caption{\textbf{Illustrations of network metrics.} A) Network density is computed for each hex grid cell. B) Network fragmentation is measured as the size and spatial distribution of disconnected components. Each color represent a separate component. C) Network reach is measured from a starting point (black node) up to a specified distance threshold. Reachable network within \num[round-mode=places,round-precision=0]{10} km (red) and \num[round-mode=places,round-precision=0]{15} km (blue). Unreachable network in gray.}
\label{fig:metric-illustrations}
\end{figure}

\textbf{Network density}, i.e. the amount of bicycle infrastructure in a given area, is a common metric for evaluating bicycle conditions \citep{schoner_missing_2014, kamel_impact_2021}. We measure first, \textit{absolute} network density as km of network length per \si{km^{2}} for each LTS in all hex grid cells, and second, \textit{relative} network density as the local percentage of each LTS out of the total network length in a grid cell (Fig.~\ref{fig:metric-illustrations}A). Measuring road network length in a data set optimized for routing and with various level of granularity in mappings of road geometries, separate bicycle tracks, etc.~is a non-trivial issue \citep{lucas-smith_is_2019}. To simplify, we compute network length as the length of all link geometries, disregarding information on number of lanes and allowed driving directions.

\textbf{Network fragmentation} is measured as the fragmentation into disconnected components for each LTS network (Fig.~\ref{fig:metric-illustrations}B). A disconnected component is a subset of a network where all nodes of the component are connected, but none of the component nodes can reach the remainder of the network. The consequence of disconnected components for low-stress infrastructure is that cyclists either are forced onto high-stress infrastructure or prevented from cycling any further, thus limiting the mobility of more risk-averse cyclists. We compute the number and length of the disconnected components and calculate the size of the largest connected component (LCC) for each LTS network. All fragmentation metrics are computed for the entire study area and locally at the hex grid level. At the grid level, disconnected components are counted as the number of different disconnected components intersecting a cell.

\textbf{Network reach} is used to capture the interplay between network density and connectivity. Network reach is  the network length that can be reached from a point moving in all possible directions along the network up to a given distance threshold \citep{peponis_connectivity_2008, feng_algorithms_2019} (Fig.~\ref{fig:metric-illustrations}C). The metric was first suggested by \cite{peponis_connectivity_2008} to capture the idea of `potential' movements in an area, arguing that it not only is important to be able to reach a given destination, but also to have access to a wider range of possible routes and locations. The idea behind network reach is consistent with research emphasizing the link between accessibility and active mobility \citep{faghih_imani_cycle_2019}. We compute network reach from each hex cell for all LTS networks for distance thresholds \num[round-mode=places,round-precision=0]{1}, \num[round-mode=places,round-precision=0]{2}, \num[round-mode=places,round-precision=0]{5}, \num[round-mode=places,round-precision=0]{10}, and \num[round-mode=places,round-precision=0]{15} km. For each LTS network, we use the largest connected component intersecting with the cell and take the node closest to the grid centroid as the starting point. 

The way in which dedicated bicycle infrastructure is mapped often leads to an artificially high fragmentation of networks of dedicated bicycle infrastructure, due to e.g.~imprecise link geometries and varying practices for mapping bicycle tracks and lanes across intersections \citep{schoner_missing_2014, viero_bikedna_2024}. For that reason, previous studies have limited the analysis to the largest connected component \citep{van_der_meer_assessment_2024} or closed gaps between cycleways under a certain length threshold \citep{schoner_missing_2014, houde_ride_2018}. We choose to close gaps of max 30 meters length between the same LTS level when computing network connectivity and reach, based on a manual assessment of LTS network gaps. The distance threshold for closing gaps does influence the exact number of closed gaps and conversely the number of disconnected components, but does not change the general patterns for network fragmentation (Fig.~\ref{SI-fig:gap-threshold}).

As a final data-preprocessing step, disconnected components that do not include dedicated bicycle infrastructure and are up to 100 meters long, or components only with the road type `track' or `footway' and length of max 500 meters are dropped, to avoid artificially high fragmentation from isolated links not part of the main road network. These components were confirmed to represent roads and paths in inaccessible private and fenced off areas, but missing a `private' tag in OSM.

\begin{figure}[H]
\centering
\includegraphics[width=0.99\textwidth]{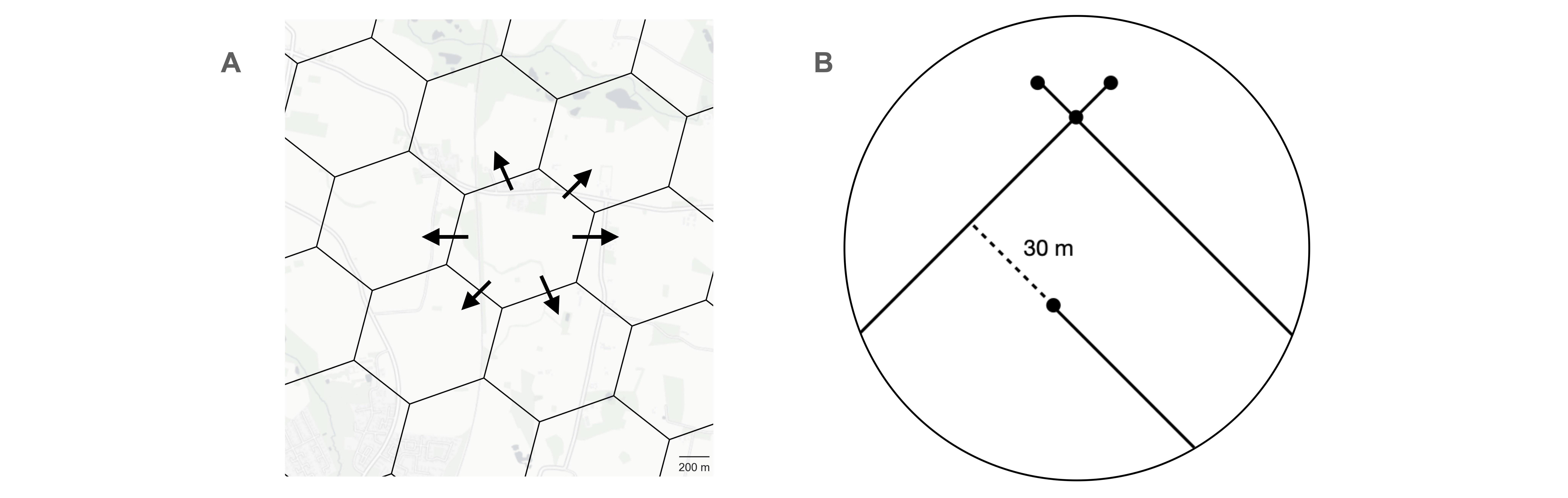}
\caption{\textbf{Data processing} A) Hex grid used for aggregation of results. The 6 nearest cells to each hex cell are considered its neighbors in the spatial weight matrix. Except for hexagons along edges of the study area, the 6 nearest neighbors will be the 6 adjacent hexagons sharing an edge. B) Gaps between network links of the same LTS level that are $\leq$ \num[round-mode=places,round-precision=0]{30} meters long are closed to compensate for artificially high fragmentation of especially dedicated bicycle infrastructure.}
\label{fig:hexgrid-illustrations}
\end{figure}

To assess whether the distribution and connectivity of low and high-stress bicycle infrastructure are spatially clustered, we analyze the degree of spatial autocorrelation for all results aggregated at the hex grid level. We use Moran's I to check for global spatial clustering, testing if similar values generally tends to be located in close proximity. To identify specific local clusters of similar and dissimilar values, we use Local Moran's I \citep{anselin_local_1995, rey_geographic_2020}. Both global and local Moran's I are computed using a spatial weight matrix based on the k-nearest neighbors with k = \num[round-mode=places,round-precision=0]{6} (Fig.~\ref{fig:hexgrid-illustrations}A). In a final identification of high and low bikeability areas, a k-means clustering algorithm is run on selected network metrics (Section \ref{subsection:bikeability-clusters}).

Data processing and analysis are implemented in PostgreSQL and Python, primarily using PostGIS \citep{postgis_psc_postgis_2023}, pgRouting \citep{pgrouting_project_pgrouting_2024}, GeoPandas \citep{jordahl_geopandasgeopandas_2021}, pysal \citep{rey_pysal_2007}, sklearn \citep{pedregosa_scikit-learn_2011}, and H3 \citep{uber_h3-py_2023}.

\section{Results}
\label{section:results}

Below, we present our findings organized in four sections: one for each of the three quality metrics (density, fragmentation, reach) and a final section for the bikeability clusters.

\subsection{Density}
\label{subsection:density}

\begin{figure}[H]
\centering
\includegraphics[width=0.975\textwidth]{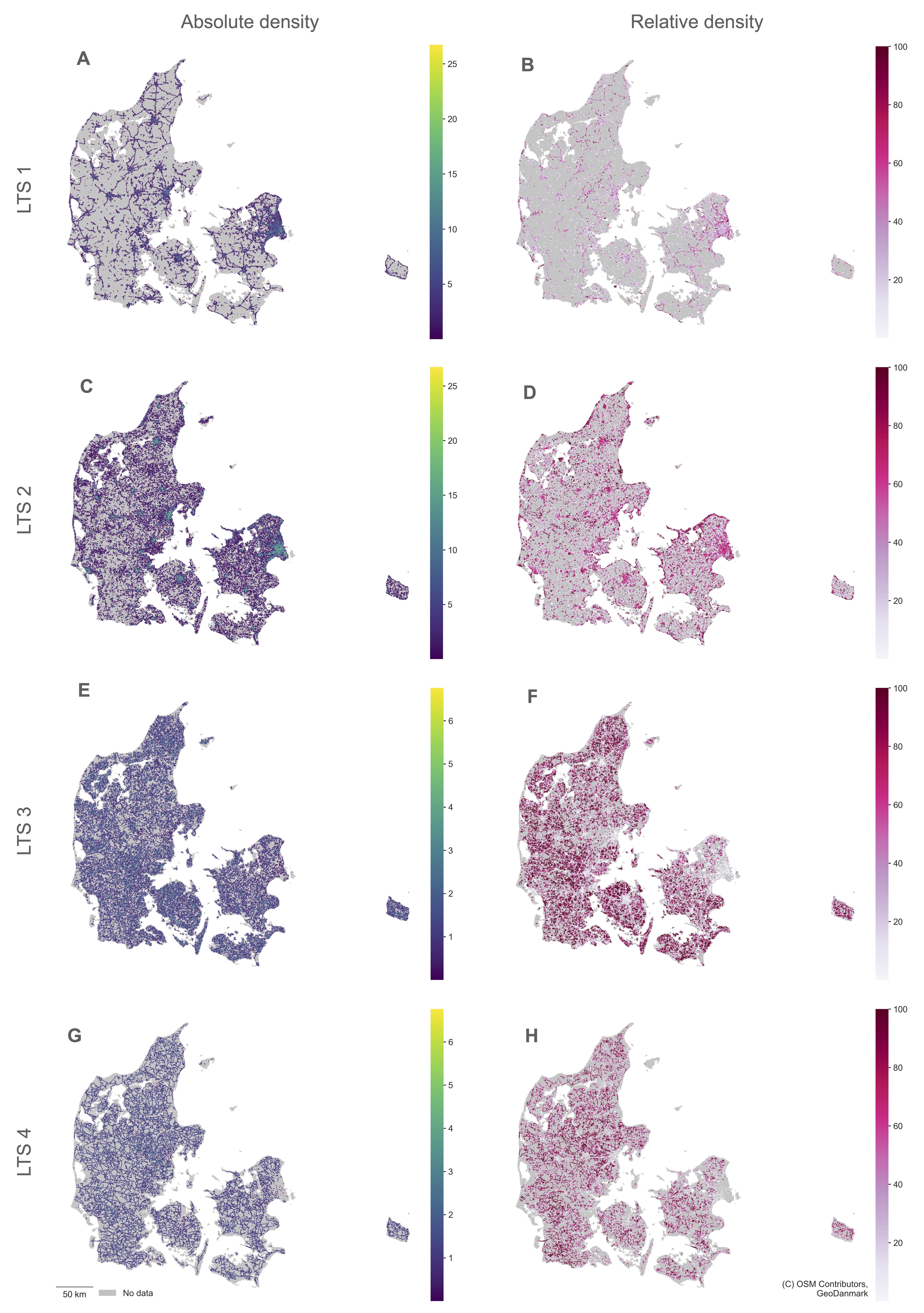}
\caption{\textbf{Network density.} A) LTS 1 -- absolute density. B) LTS 1 -- relative density. C) LTS 2 -- absolute density. D) LTS 2 -- relative density. C) LTS 3 -- absolute density.  D) LTS 3 -- relative density. E) LTS 4 -- absolute density. F) LTS 4 -- relative density. Values for absolute densities are in km/\si{km^{2}}, relative densities are \%. Network densities for low-stress infrastructure are spatially clustered, with high values in population centers. Values for LTS 3 and 4 infrastructure are more spatially dispersed. Observe that the color bars have different value ranges. See Fig.~\ref{SI-fig:car-total-density} for car and total network densities.}
\label{fig:density}
\end{figure}

First, we measure network density as the \textit{absolute} local density (km/\si{km^{2}}) of each LTS level, the car network, and the total network at the hex grid level. Second, we measure the \textit{relative} density as the local \textit{percentage} of each network type out of the length of the total network in a grid cell. For the entire study area, the total network contains almost the same amount of very low and very high-stress infrastructure (LTS 1: \num[round-mode=places,round-precision=0]{15.49}\%, LTS 4: \num[round-mode=places,round-precision=0]{15.65}\%). The combined low-stress infrastructure (LTS 1 and 2) constitutes more than half of the total network length ($\sim$\num[round-mode=places,round-precision=0]{53}\%) (Fig.~\ref{fig:lts-share}A and Table \ref{table:lts-overview}). Denmark thus appears to have a substantial amount of low-stress infrastructure  (although it is difficult to compare LTS distributions to other countries, as most other studies only include urban areas). At the local scale (Fig.~\ref{fig:density}), we, however, observe significant differences both in the absolute and relative amount of low and high-stress infrastructure:

\begin{itemize}
    \item Absolute LTS 1 density is highest around highly populated areas (Fig.~\ref{fig:density}A). Relative LTS 1 density tends to be high in densely populated areas, in some spatially concentrated areas along the coast, and along some rural roads connecting different towns and cities (Fig.~\ref{fig:density}B).
    \item LTS 2 infrastructure shows a much larger spatial dispersion than LTS 1 in both absolute and relative density, but is also highest in areas with high population densities (Fig.~\ref{fig:density}C,D).
    \item LTS 3 infrastructure is found almost across the entire country, but makes up a relatively low share of the network in highly populated areas (Fig.~\ref{fig:density}E,F).
    \item LTS 4 infrastructure is similarly dispersed across the study area, although more spatially concentrated than LTS 3, and is predominantly \textit{absent} in the most populated areas (Fig.~\ref{fig:density}G,H) (see Fig.~\ref{SI-fig:dens-pop-corr} for correlations between network and population densities).
\end{itemize}

Thus, both absolute and relative LTS 1 and 2 densities tend to be highest in areas with a high population and total network density. Low-stress infrastructure also tends to be spatially concentrated in relatively small areas, resulting in high maximum density values for the low-stress networks, despite LTS 1 and 2 rarely making out the majority of the local infrastructure (Table \ref{table:density-overview}). Comparison of absolute and relative network density maps reveal that areas with high \textit{low} stress densities tend to have fairly low \textit{high} stress densities, and vice versa (Fig.~\ref{fig:density} and Fig.~\ref{SI-fig:cluster-kde}). A test for spatial autocorrelation confirms that all LTS levels, the car network, and the total network density are spatially clustered, although to a smaller extent for LTS 3 and 4 (Table \ref{SI-table:moransi}, Fig.~\ref{SI-fig:lisa-dens} - \ref{SI-fig:lisa-relative}).

\begin{table}[H]
\centering
\begin{tabular}{crrrrrrrr}
\toprule
\textbf{Network} & \textbf{Min} & \textbf{Mean} & \textbf{Median} & \textbf{Max} & \textbf{Min} & \textbf{Mean} & \textbf{Median} & \textbf{Max} \\
\textbf{level} & \textbf{density} & \textbf{density} & \textbf{density} & \textbf{density} & \textbf{share} & \textbf{share} & \textbf{share} & \textbf{share} \\
\midrule
LTS 1 & \num[round-mode=places,round-precision=0]{0.0} & \num{2.03} & \num{1.34} & \num{18.64} & \num[round-mode=places,round-precision=0]{0.0}\% & \num[round-mode=places,round-precision=0]{31.0}\% & \num[round-mode=places,round-precision=0]{27.5}\% & \num[round-mode=places,round-precision=0]{100.0}\% \\
LTS 2 & \num[round-mode=places,round-precision=0]{0.0} & \num{2.14} & \num{0.85} & \num{28.72} & \num[round-mode=places,round-precision=0]{0.0}\% & \num[round-mode=places,round-precision=0]{42.9}\% & \num[round-mode=places,round-precision=0]{36.9}\% & \num[round-mode=places,round-precision=0]{100.0}\% \\
LTS 3 & \num[round-mode=places,round-precision=0]{0.0} & \num{1.19} & \num{1.15} & \num{6.54} & \num[round-mode=places,round-precision=0]{0.0}\% & \num[round-mode=places,round-precision=0]{55.3}\% & \num[round-mode=places,round-precision=0]{51.8}\% & \num[round-mode=places,round-precision=0]{100.0}\% \\
LTS 4 & \num[round-mode=places,round-precision=0]{0.0} & \num{1.17} & \num{1.21}  & \num{6.76} & \num[round-mode=places,round-precision=0]{0.0}\% & \num[round-mode=places,round-precision=0]{52.9}\% & \num[round-mode=places,round-precision=0]{50.0}\% & \num[round-mode=places,round-precision=0]{100.0}\% \\
Car & \num[round-mode=places,round-precision=0]{0.0} & \num{3.04} & \num{2.09} & \num{34.67} & \num[round-mode=places,round-precision=0]{0.3}\%  & \num[round-mode=places,round-precision=0]{94.9}\% & \num[round-mode=places,round-precision=0]{100.0}\% & \num[round-mode=places,round-precision=0]{100.0}\% \\
Total network & \num[round-mode=places,round-precision=0]{0.0} & \num{3.39} & \num{2.17} & \num{43.51} & \num[round-mode=places,round-precision=0]{100.0}\% & \num[round-mode=places,round-precision=0]{100.0}\%  & \num[round-mode=places,round-precision=0]{100.0}\%  & \num[round-mode=places,round-precision=0]{100.0}\% \\
\bottomrule
\end{tabular}
\smallskip
\caption{\textbf{Network densities}. Values for absolute densities are in km/\si{km^{2}}. High-stress infrastructure (LTS 3 and 4) are more spatially dispersed, contributing to lower values for absolute maximum density.}
\label{table:density-overview}
\end{table}

\subsection{Fragmentation}
\label{subsection:fragmentation}

\begin{figure}[H]
    \centering
    \includegraphics[width=0.99\linewidth]{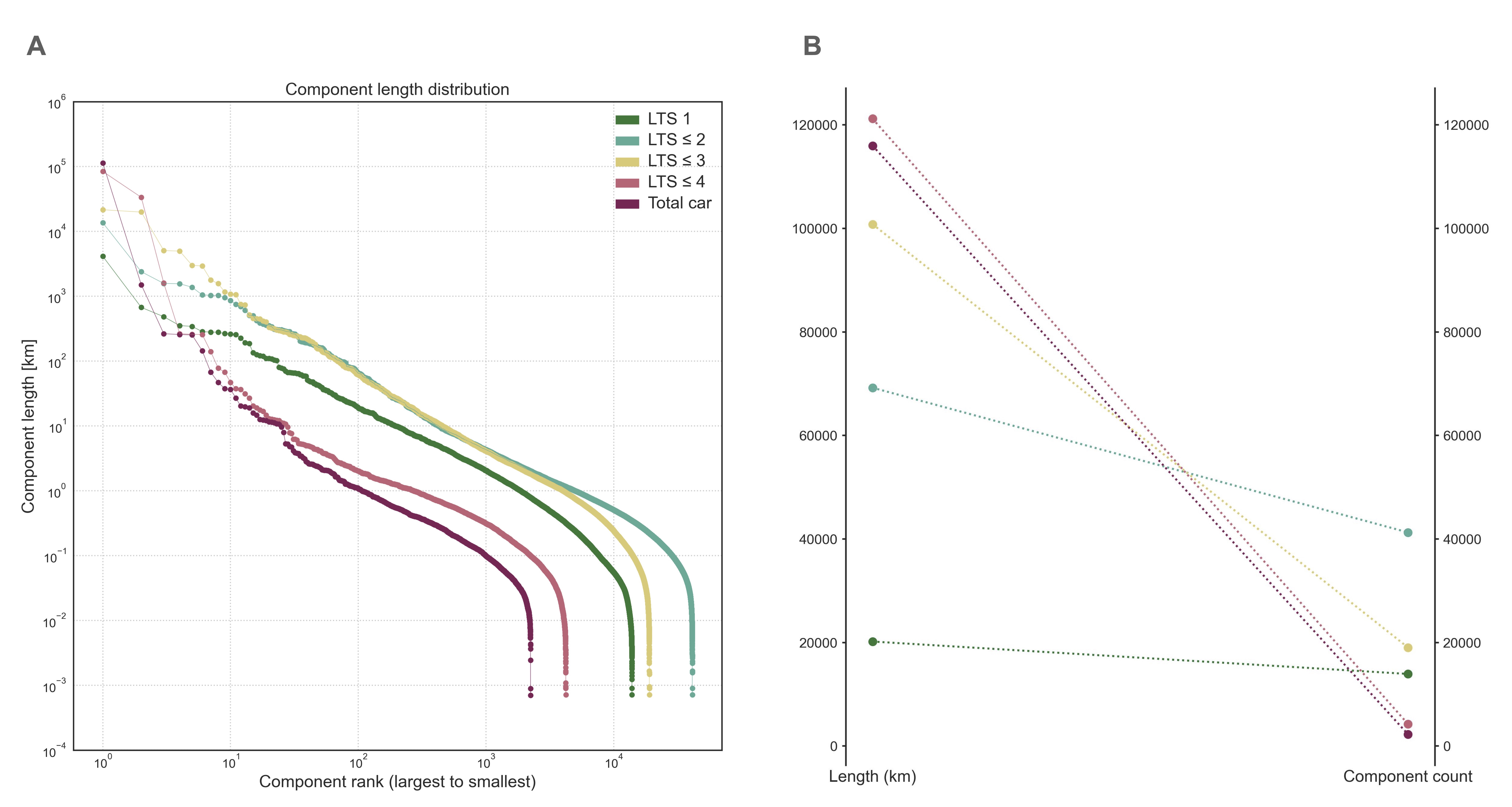}
    \caption{\textbf{Component size ranking.} A) Zipf plot ranking the length of components grouped by LTS level in descending order on a log-log scale. B) Comparison of network length and component count. The component count increases when moving from LTS 1 to LTS$\leq$2. The network fragmentation decreases once LTS 3 is included, but only falls substantially when LTS 4 infrastructure is considered.}
    \label{fig:zipf-length-vs-comp}
\end{figure}

We measure network fragmentation as the total number of disconnected components, count the number of components at the hex grid level, and compute the size of the largest connected component (LCC) for all network levels (Fig.~\ref{fig:metric-illustrations}B). The LTS 1 and LTS$\leq$2 networks show  high  fragmentation, evident in the high number of disconnected components (\num[round-mode=places,round-precision=0]{13903} and \num[round-mode=places,round-precision=0]{41247}), the small mean component size (< \num[round-mode=places,round-precision=0]{2} km), and their LCC forming a relatively low share of the total network length ($\sim$\num[round-mode=places,round-precision=0]{20}\%) (Table \ref{table:components}). The larger fragmentation of the lower stress networks is also illustrated on Fig.~\ref{fig:zipf-length-vs-comp}A, which ranks components by their size for all network levels. The component size rankings are mostly similar for the total network, the car network, and the LTS$\leq$4 network. These three networks have a few very large components (the leftmost markers between rank \si{10^0} and \si{10^1}) and relatively few smaller components. The component size rankings for LTS 1, $\leq$2, and $\leq$3 follow very different patterns with flatter curves that reach higher component ranks, especially for the LTS$\leq$2 network, which has many very small disconnected components (<\num[round-mode=places,round-precision=0]{1}km, or <\si{10^3} on Fig.~\ref{fig:zipf-length-vs-comp}A).

\begin{table}[H]
\centering
\begin{tabular}{lrrrr}
\toprule
\textbf{Network level}& \textbf{Comp. count} &\textbf{Mean comp. size} &\textbf{LCC size} &\textbf{Share of LCC} \\
\midrule
LTS 1 & \num[round-mode=places,round-precision=0]{13903} & \num{1.45} km  & \num[round-mode=places,round-precision=0]{4105.77} km & \num[round-mode=places,round-precision=0]{20.36} \%\\
LTS$\leq$2 & \num[round-mode=places,round-precision=0]{41247} & \num{1.68} km & \num[round-mode=places,round-precision=0]{13507.00} km & \num[round-mode=places,round-precision=0]{19.52} \%\\
LTS$\leq$3 & \num[round-mode=places,round-precision=0]{19014} & \num{5.30} km & \num[round-mode=places,round-precision=0]{21462.91} km & \num[round-mode=places,round-precision=0]{21.30} \%\\
LTS$\leq$4 & \num[round-mode=places,round-precision=0]{4220} & \num{28.72} km & \num[round-mode=places,round-precision=0]{83577.29} km & \num[round-mode=places,round-precision=0]{68.98} \%\\
Car & \num[round-mode=places,round-precision=0]{2235} & \num{51.9} km & \num[round-mode=places,round-precision=0]{112693.39} km & \num[round-mode=places,round-precision=0]{97.19} \%\\
Total & \num[round-mode=places,round-precision=0]{2331} & \num{55.88}  km & \num[round-mode=places,round-precision=0]{126594.14} km & \num[round-mode=places,round-precision=0]{97.22} \%\\
\bottomrule
\end{tabular}
\smallskip
\caption{\textbf{Network fragmentation}. `Share of LCC' represents the share of the total length of each LTS network represented by the largest connected component (LCC). The high fragmentation of lower-stress infrastructure results in a high component count, small mean component size, and the LCC only making up a low share of the total length.}
\label{table:components}
\end{table}

Importantly, despite the drastic differences for the country as a whole, the mean local component count at the hex grid scale is only slightly higher for the low-stress networks (compare \num[round-mode=places,round-precision=2]{1.89} to \num[round-mode=places,round-precision=2]{1.35} in Table \ref{table:local-component-count}, and see Fig.~\ref{SI-fig:component-rug}). The high \textit{total} component count but relatively low \textit{local} component count for LTS 1 and LTS$\leq$2 indicate that low-stress infrastructure is well connected \textit{locally}, but that areas with low-stress infrastructure form isolated `islands' (Fig.~\ref{fig:lts-subgraphs}A-D). The number of disconnected components for the LTS$\leq$2 network is more than double the number of components when including only LTS 1. That the number of components does not decrease when including LTS 2 indicates that LTS 2 infrastructure generally does not connect otherwise disconnected LTS 1 components but instead forms new islands of low-stress infrastructure. The component count drops significantly when including LTS 3 infrastructure, as evident from the steeper slope for LTS$\leq$3, but only when LTS 4 infrastructure is included does it reach levels comparable to the connectivity for cars (Fig.~\ref{fig:zipf-length-vs-comp}B and Table \ref{table:components}). This pattern matches the findings of previous research, which similarly found that it is necessary to include LTS 4 infrastructure to substantially decrease network fragmentation \citep{crist_fear_2019, reggiani_multi-city_2023}.

\begin{table}[H]
\centering
\begin{tabular}{lrrrr}
\toprule
\textbf{Network level}& \textbf{Min} &\textbf{Mean} &\textbf{Median} &\textbf{Max} \\
\midrule
LTS 1 & \num[round-mode=places,round-precision=0]{1} & \num{1.83} & \num{1} & \num[round-mode=places,round-precision=0]{32} \\
LTS$\leq$2 & \num[round-mode=places,round-precision=0]{1} & \num{1.89} & \num{1} & \num[round-mode=places,round-precision=0]{26} \\
LTS$\leq$3 & \num[round-mode=places,round-precision=0]{1} & \num{1.35} & \num{1} & \num[round-mode=places,round-precision=0]{18} \\
LTS$\leq$4 & \num[round-mode=places,round-precision=0]{1} & \num{1.08} & \num{1} & \num[round-mode=places,round-precision=0]{18} \\
Car & \num[round-mode=places,round-precision=0]{1} & \num{1.04} & \num{1} & \num[round-mode=places,round-precision=0]{20} \\
Total & \num[round-mode=places,round-precision=0]{1} & \num{1.04} & \num{1} & \num[round-mode=places,round-precision=0]{18}\\
\bottomrule
\end{tabular}
\smallskip
\caption{\textbf{Local component count}. The number of disconnected components for each network level at the hex grid level. The local component count is generally low, but slightly higher for LTS 1, LTS$\leq$2, and LTS$\leq$3.}
\label{table:local-component-count}
\end{table}

\begin{figure}[H]
\centering
\includegraphics[width=0.99\textwidth]{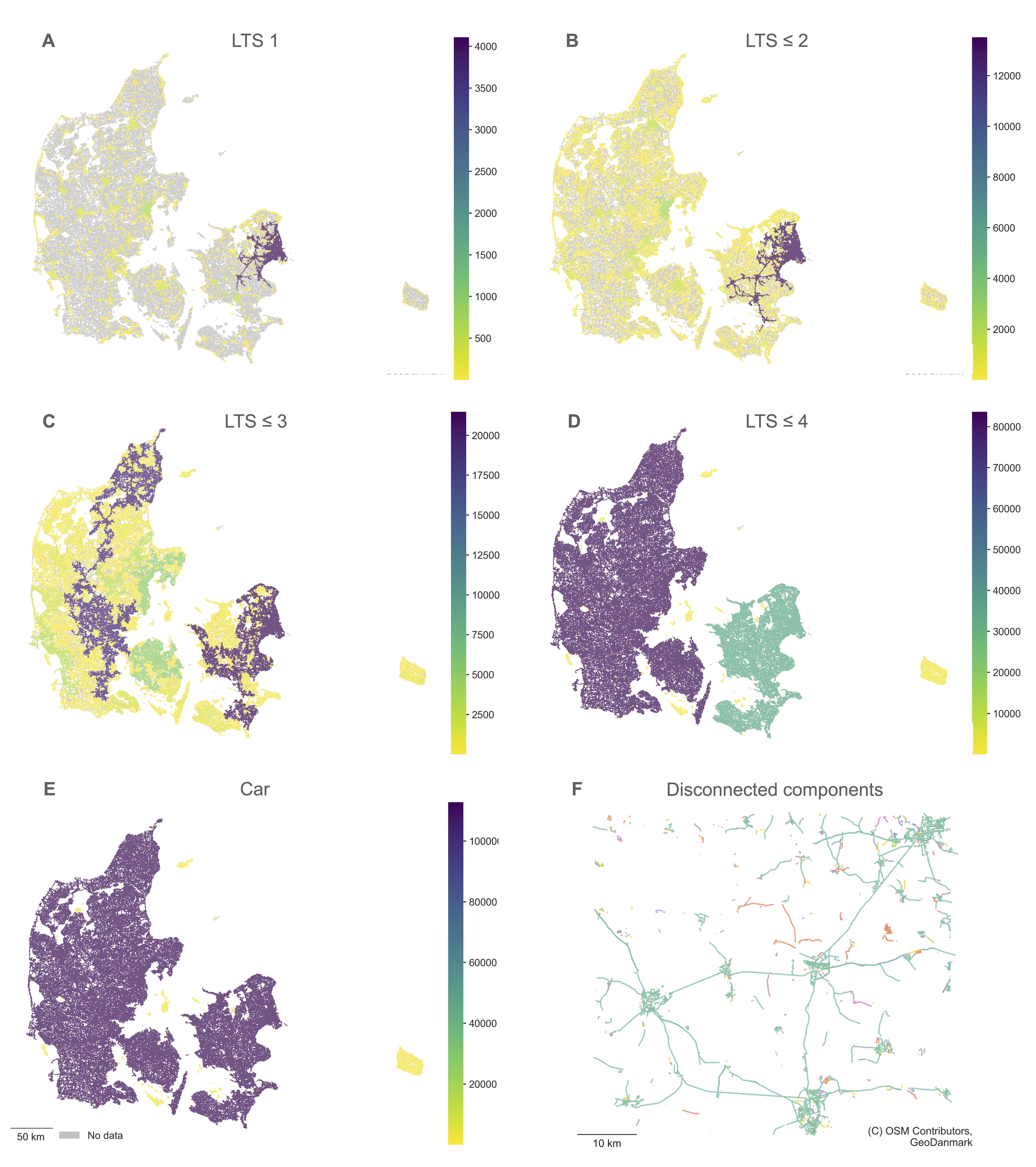}
\caption{\textbf{Largest component length.} Length of the LCC (km) at the hex grid level. A) LTS 1. B) LTS$\leq$2. C) LTS$\leq$3. D) LTS$\leq$4. E) Car. F) Illustration of disconnected components for LTS 1. The sizes of the LCCs for lower stress infrastructure are spatially clustered, with larger LCCs in highly populated areas. For the LTS$\leq$4 and car networks, the size of the local LCC is a question of landscape topography.}
\label{fig:largest-component-length}
\end{figure}

To understand whether the networks are more fragmented in some areas than others, we compute the spatial autocorrelation of the local component count. The spatial autocorrelation for components per km at the hex grid scale is not statistically significant (Table \ref{SI-table:moransi} and Fig.~\ref{SI-fig:lisa-largest-comp}). The lack of evidence of particularly high fragmentation in certain areas again indicates that the fragmentation of low-stress infrastructure does not happen at the very local scale. Instead, fragmentation occurs due to the spatial concentration of areas with or without any low-stress infrastructure. In contrast, the size of the local LCC for low-stress infrastructure is spatially clustered, with the most populated areas having the largest local components, particularly for low-stress infrastructure (Fig.~\ref{fig:largest-component-length}). The very uneven distribution of low-stress infrastructure thus results in a road network that, for cyclists unable or unwilling to bike on high-stress roads, is highly fragmented across larger distances outside of highly populated areas, but well connected locally.

\subsection{Reach}
\label{subsection:reach}

We measure network reach as the reachable network when moving in all possible directions along the network up to a maximum distance threshold (Fig.~\ref{fig:metric-illustrations}C). We compare network reach for each network level at different distance thresholds (\num[round-mode=places,round-precision=0]{1}, \num[round-mode=places,round-precision=0]{2}, \num[round-mode=places,round-precision=0]{5}, \num[round-mode=places,round-precision=0]{10}, and \num[round-mode=places,round-precision=0]{15} km), and compare bicycle network reach for each LTS network with the car reach in each hex cell. The mean network reach is lowest for the LTS 1 network (\num[round-mode=places,round-precision=1]{23.3} km), with a median reach of \num[round-mode=places,round-precision=1]{7.9} km (Table \ref{table:reach-overview}).  Due to its large fragmentation, the LTS$\leq$ 2 network has the lowest median reach (\num[round-mode=places,round-precision=1]{2.7} km), despite having a mean reach of \num[round-mode=places,round-precision=1]{42.9} km and being more than twice the length of LTS 1 (Table \ref{table:lts-overview}). The median network reach increases substantially to \num[round-mode=places,round-precision=1]{26.7} km when LTS 3 infrastructure is introduced. The LTS$\leq$4 network surpasses the car network in mean (\num[round-mode=places,round-precision=1]{100.50} km vs. \num[round-mode=places,round-precision=1]{93.30} km) and median (\num[round-mode=places,round-precision=1]{76.5} km vs. \num[round-mode=places,round-precision=1]{75.2} km) reach, as the LTS$\leq$4 network includes both most of the car network and bicycle paths closed to car traffic. Remarkably, the maximum reach for LTS$\leq$2, $\leq$3, $\leq$4, and car network are all around \num[round-mode=places,round-precision=0]{1100} km. The comparable maximum reach values reveal that there are locations that have the necessary preconditions for a high low-stress reach, namely a high low stress density \textit{and} good low-stress connectivity.

\begin{table}[H]
\centering
\begin{tabular}{crrrr}
\toprule
\textbf{Network level} & \textbf{Mean reach} & \textbf{Median reach} & \textbf{Max reach} & \textbf{Mean \% of car reach} \\
\midrule
LTS 1 & \num[round-mode=places,round-precision=1]{23.3} km & \num[round-mode=places,round-precision=1]{7.9} km & \num[round-mode=places,round-precision=1]{311.00} km & \num[round-mode=places,round-precision=0]{18.9} \%\\
LTS$\leq$2 & \num[round-mode=places,round-precision=1]{42.9} km & \num[round-mode=places,round-precision=1]{2.7} km & \num[round-mode=places,round-precision=1]{1089.04} km & \num[round-mode=places,round-precision=0]{36.5} \%\\
LTS$\leq$3 & \num[round-mode=places,round-precision=1]{57.0} km & \num[round-mode=places,round-precision=1]{26.7} km & \num[round-mode=places,round-precision=1]{1136.97} km & \num[round-mode=places,round-precision=0]{59.0} \%\\
LTS$\leq$4 & \num[round-mode=places,round-precision=1]{100.5} km & \num[round-mode=places,round-precision=1]{76.5} km & \num[round-mode=places,round-precision=1]{1138.35} km & \num[round-mode=places,round-precision=0]{118.8} \%\\
Car & \num[round-mode=places,round-precision=1]{93.3} km & \num[round-mode=places,round-precision=1]{75.2} km & \num[round-mode=places,round-precision=1]{975.36} km & - \\
\bottomrule
\end{tabular}
\smallskip
\caption{\textbf{Network reach}. Network reach at different network levels for distance threshold = \num[round-mode=places,round-precision=0]{5} km.}
\label{table:reach-overview}
\end{table}

\begin{figure}[H]
    \centering
\includegraphics[width=0.99\linewidth]{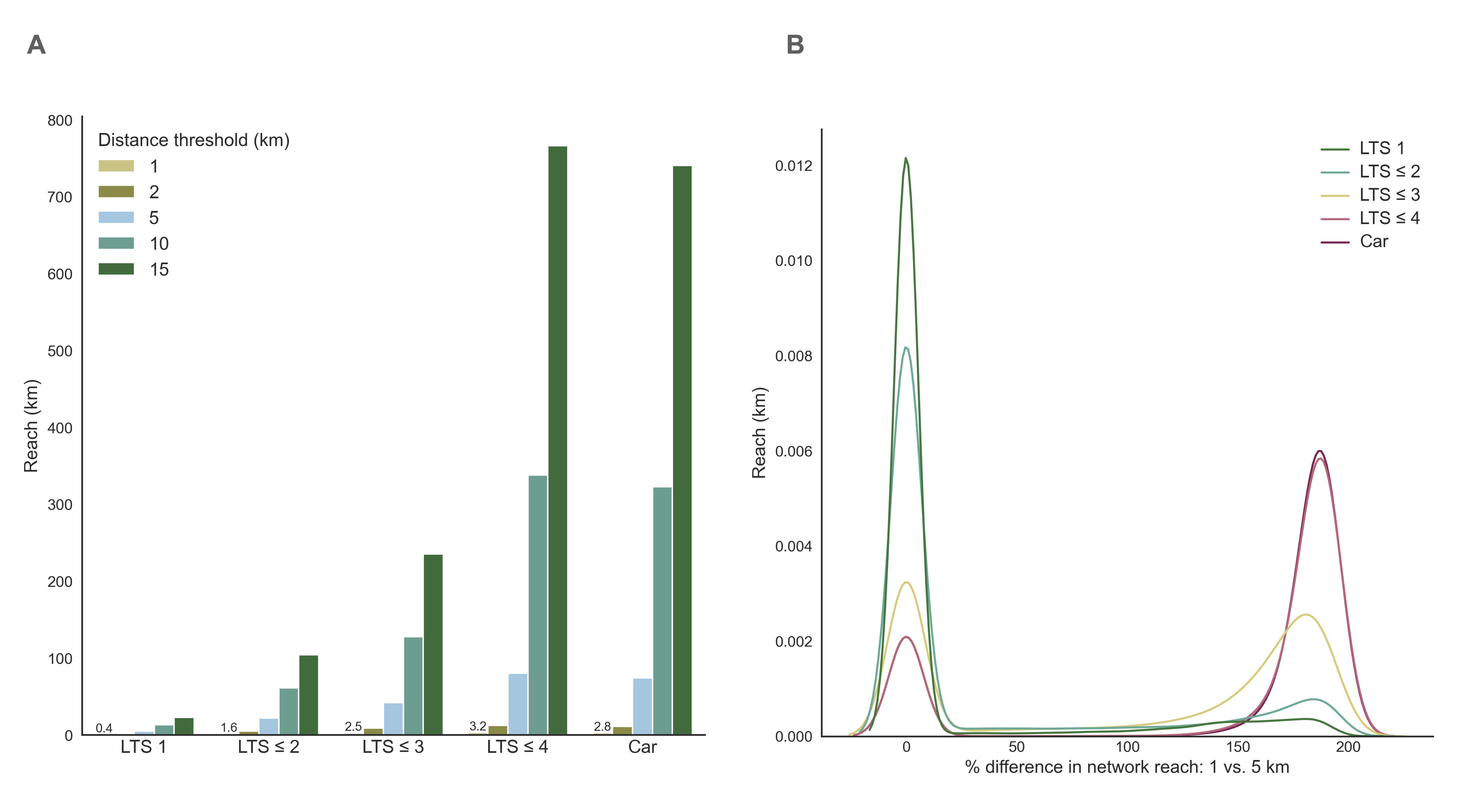}
    \caption{\textbf{Comparison of network reach.} A)  Mean network reach for each network level at distance thresholds \num[round-mode=places,round-precision=0]{1}, \num[round-mode=places,round-precision=0]{2}, \num[round-mode=places,round-precision=0]{5}, \num[round-mode=places,round-precision=0]{10}, and \num[round-mode=places,round-precision=0]{15} km. B) KDE plot comparing network reach between \num[round-mode=places,round-precision=0]{1} and \num[round-mode=places,round-precision=0]{5} km distance thresholds. Many locations have no reach improvement for the low-stress network when increasing the distance threshold from \num[round-mode=places,round-precision=0]{1} to \num[round-mode=places,round-precision=0]{5} km (\num[round-mode=places,round-precision=0]{0}\% increase in reach). Conversely, when using the high-stress networks almost all locations see reach improvements when increasing the distance threshold.}
    \label{fig:reach-dist-comparison}
\end{figure}

Increasing distance thresholds allows longer bicycle trips, which naturally leads to a larger network reach (Fig.~\ref{fig:reach-dist-comparison}A). However, for a large number of locations, there is no reach improvement as the distance threshold increases, due to the high fragmentation and spatial clustering of low-stress infrastructure that only allows for very short low-stress bicycle trips. In Fig.~\ref{fig:reach-dist-comparison}B, we therefore see a high probability of no improvement in reach for low stress networks when increasing the distance threshold from \num[round-mode=places,round-precision=0]{1} to \num[round-mode=places,round-precision=0]{5} km, as evident from the high density around 0\% improvement for LTS 1 and LTS$\leq$2. This lack of reach improvement when increasing distance thresholds rarely occurs for networks including higher-stress infrastructure (evident from a low density for 0\% increase for LTS$\leq$3 and LTS$\leq$4 and high density for \num[round-mode=places,round-precision=0]{150}+\% reach increase) (Fig.~\ref{fig:reach-dist-comparison}B). Even so, there do exists locations with a large growth in low-stress network reach as the distance threshold increases.

\begin{figure}[H]
\centering
\includegraphics[width=0.99\textwidth]{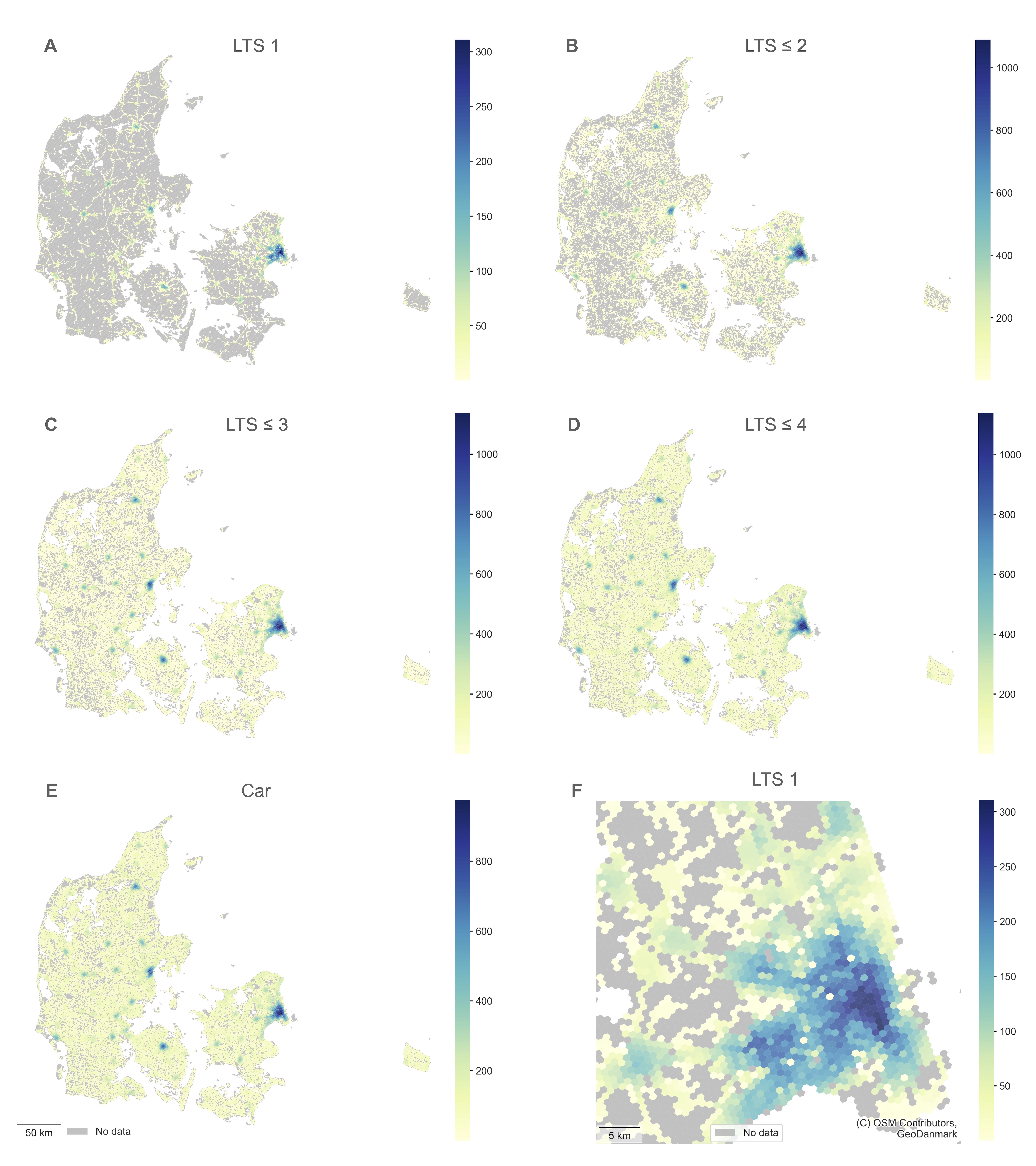}
\caption{\textbf{Network reach.} Network reach (km) within a \num[round-mode=places,round-precision=0]{5} km distance threshold. A) LTS 1. B) LTS$\leq$2. C) LTS$\leq$3. D) LTS$\leq$4. E) Car. F) Detail map of LTS 1 network reach in the Greater Copenhagen area.}
\label{fig:reach}
\end{figure}

\begin{figure}[H]
\centering
\includegraphics[width=0.99\textwidth]{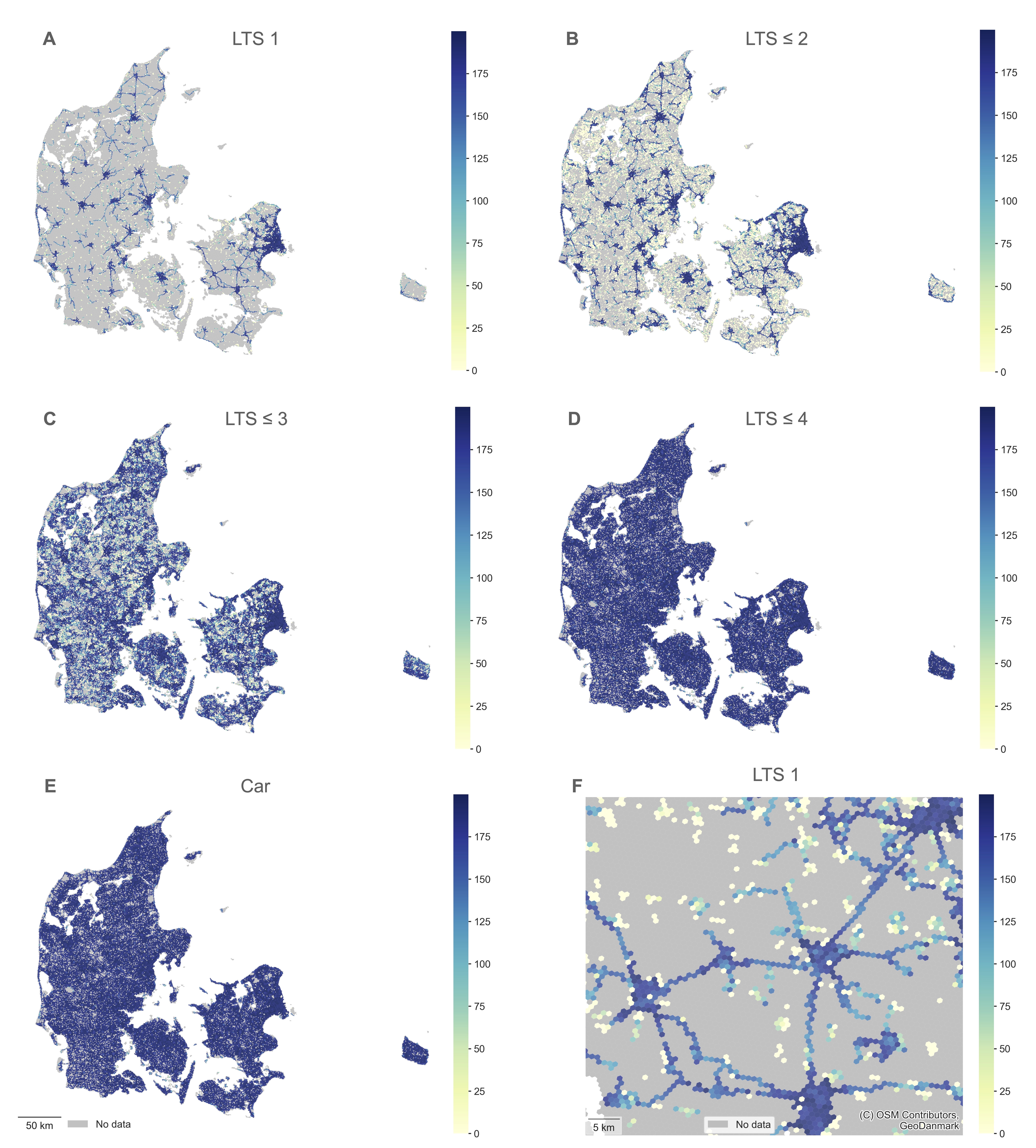}
\caption{\textbf{Network reach increase with increased distance threshold.} Percent increase in network reach  between \num[round-mode=places,round-precision=0]{1} and \num[round-mode=places,round-precision=0]{5} km distance thresholds. A) LTS 1. B) LTS$\leq$2. C) LTS$\leq$3. D) LTS$\leq$4. E) Car. F) Detail map of network reach increases for LTS 1. While absolute network reach is highly correlated with network density, reach increases for longer distance thresholds also happen in low-density but well connected locations.}
\label{fig:reach-increase}
\end{figure}

The network reach is significantly spatially clustered (Table \ref{SI-table:moransi} and Fig.~\ref{SI-fig:lisa-reach}), with high reach values in areas with high network densities (Fig.~\ref{fig:reach}). Therefore, high network reach, particularly in the lower stress networks, is also correlated with high population densities (Figs.~\ref{SI-fig:dens-reach-corr} and \ref{SI-fig:hex-corr-heatmap}). Furthermore, areas with protected bicycle tracks along main roads connecting urban centers also have large \textit{increases} in network reach when increasing distance thresholds, despite modest low-stress densities (Fig.~\ref{fig:reach-increase}); illustrating the effect of initiatives aimed at connecting urban, suburban, and rural areas with direct and high quality bicycle paths, such as cycle highways \citep{skov-petersen_effects_2017, filho_cycle_2024}.

In most locations outside of the largest population centers, the results for network reach, nevertheless, illustrate the effects of high fragmentation of low-stress networks: often, a location will have some low-stress infrastructure, but this low-stress infrastructure only allows for biking locally and within a short radius before forcing cyclists into mixed traffic on high-stress roads.

\subsection{Bikeability clusters}
\label{subsection:bikeability-clusters}

We have established that, nationwide, Denmark has a relatively large share of low-stress infrastructure. However, this infrastructure  is very unevenly distributed. The spatial concentration of low-stress infrastructure, combined with a high network fragmentation, results in very low network reach on the low-stress network outside of the most highly populated areas. By contrast, we also find that bicycle tracks along some rural roads result in good long-distance network reach for some low-density areas, and that when including the full bikeable network, bicycle reach is generally comparable to, if not larger than, network reach for cars. 

To condense the multi-dimensional analysis and better understand the interaction between different network metrics and their spatial distribution, we run a k-means clustering algorithm on the network metrics aggregated at the hex grid scale. Many of the network metrics are highly correlated at the local level (Fig.~\ref{SI-fig:hex-corr-heatmap}), as is often the case for network analysis metrics \citep{dill_measuring_2004}. To improve interpretability, the most correlated variables are excluded from the clustering. In the end, the included variables are: network density for each network level; percentage of local network with car traffic; network reach; and network reach increases between \num[round-mode=places,round-precision=0]{1} and \num[round-mode=places,round-precision=0]{5} km and between \num[round-mode=places,round-precision=0]{5} and \num[round-mode=places,round-precision=0]{10} km. The component count at the hex grid scale is highly correlated with network density and is therefore not included. Using the elbow method \citep{shi_quantitative_2021}, we set the number of clusters k = \num[round-mode=places,round-precision=0]{5}. Assuming that high densities of low-stress infrastructure and high network reach are signs of high bikeability, the bikeability clusters are ranked from 1 to 5 with 1 for the lowest and 5 for highest bikeability.

\bigskip

\noindent Based on the cluster means (Fig.~\ref{SI-fig:cluster-kde} and Table \ref{SI-table:cluster-means}), we summarize the clusters as:

\begin{itemize}
    \item \textbf{Cluster 1 -- High stress:} High network densities for the high-stress networks, low network densities for the low-stress networks, very low total network density, and low network reach. 
    \item \textbf{Cluster 2 -- Local low stress connectivity:} High network densities for high-stress network, low densities for low-stress network, low total network density, low network reach. Relatively high reach increases between distance threshold \num[round-mode=places,round-precision=0]{1} and \num[round-mode=places,round-precision=0]{5} km for the LTS 1, $\leq$3, and $\leq$4 networks indicating fairly good connectivity at the local scale. 
    \item \textbf{Cluster 3 -- Regional low stress connectivity:} Very similar to cluster 2, but relatively high network increases for all network levels both between distance thresholds \num[round-mode=places,round-precision=0]{1} to \num[round-mode=places,round-precision=0]{5} km and \num[round-mode=places,round-precision=0]{5} to \num[round-mode=places,round-precision=0]{10} km, indicating connections beyond the local scale. 
     \item \textbf{Cluster 4 -- High bikeability:} Low network densities for high-stress networks, high network densities for low-stress networks, high total network density, and medium network reach. 
    \item \textbf{Cluster 5 -- Highest bikeability and high density:} High densities for the low-stress networks, low densities for high-stress networks, very high total network density, and high reach and reach increases (although slightly lower reach increases for the LTS$\leq$4 and car network, compared to cluster 4). 
\end{itemize}

\begin{figure}[H]
\centering
\includegraphics[width=0.99\textwidth]{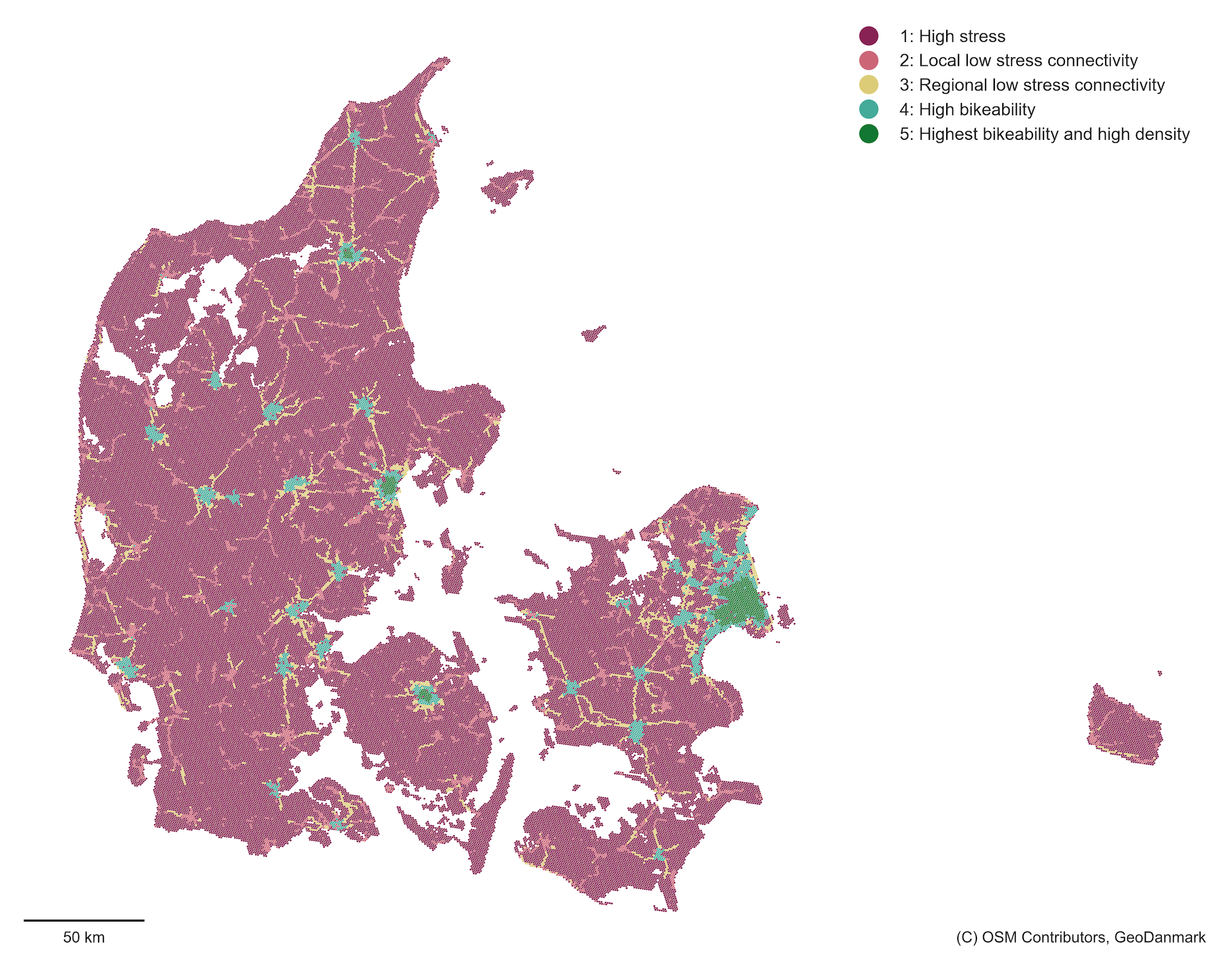}
\caption{\textbf{Bikeability clusters.} The clusters form clear spatial patterns based on urban areas and population densities despite using a clustering algorithm with no spatial constrains and not using population density as input.}
\label{fig:bikeability-clusters}
\end{figure}

\begin{figure}[H]
\centering
\includegraphics[width=0.65\textwidth]{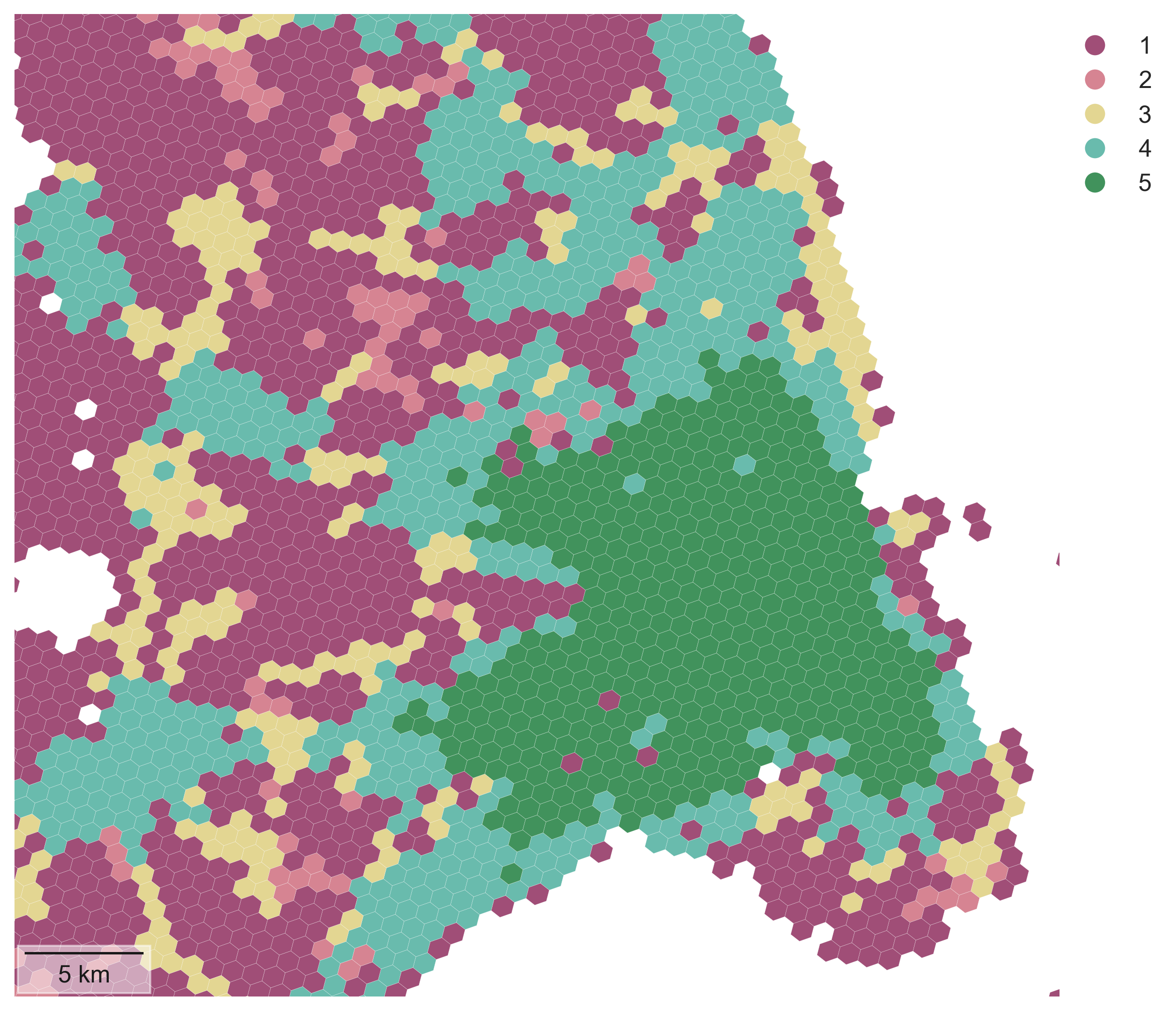}
\caption{\textbf{Detail map of bikeability clusters.} Bikeability clusters in the Greater Copenhagen area.}
\label{fig:bikeability-clusters-zoom}
\end{figure}

\begin{figure}[H]
\centering
\includegraphics[width=0.99\textwidth]{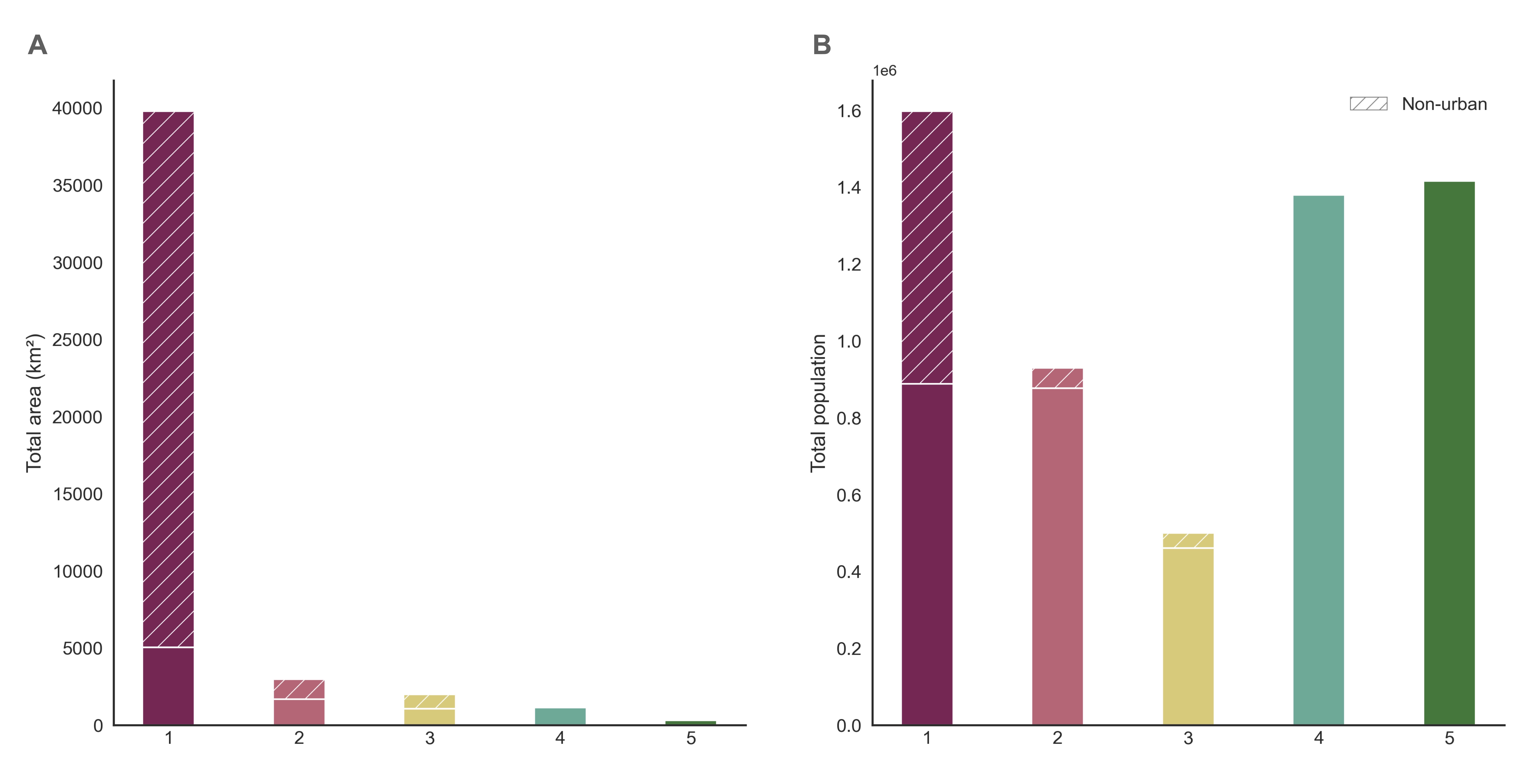}
\caption{\textbf{Total area and population in each cluster.} A) Total area. B) Total population. Although cluster 1 covers the vast majority of the study area, it only contains \num{27.417}\% of the population.}
\label{fig:cluster-area-pop}
\end{figure}

The k-means clustering method does not include any spatial constraints or information on location or spatial organization, but the resulting clusters form distinct spatial patterns (Fig.~\ref{fig:bikeability-clusters}-\ref{fig:bikeability-clusters-zoom}). Cluster 1 covers mainly non-urban areas with low population densities (Fig.~\ref{fig:cluster-area-pop}); despite covering \num[round-mode=places,round-precision=1]{85.88}\% of the country, cluster 1 only contains \num[round-mode=places,round-precision=1]{27.42}\% of the total population. Cluster 2 is the second largest cluster area-wise (\num[round-mode=places,round-precision=1]{6.44}\%), although much smaller than cluster 1, and is predominantly found in smaller towns with low population densities (\num[round-mode=places,round-precision=1]{15.96}\% of the population). Cluster 3 has \num[round-mode=places,round-precision=1]{4.34}\% of the area, is the smallest cluster population-wise (\num[round-mode=places,round-precision=1]{8.59}\%), and is mostly found in low urban, low population areas with good low-stress cycling connections to higher density areas.  Cluster 4 is geographically small (\num[round-mode=places,round-precision=1]{2.61}\% of the total area), covers mainly suburbs and medium-sized towns, and contains \num[round-mode=places,round-precision=1]{23.73}\% of the population. Finally, cluster 5 only covers \num[round-mode=places,round-precision=1]{0.72}\% of the area but has \num[round-mode=places,round-precision=1]{24.31}\% of the population. Cluster 5 is spatially concentrated and found exclusively in and around the four largest cities in Denmark (Fig.~\ref{fig:bikeability-clusters}-\ref{fig:bikeability-clusters-zoom}).

The location and extent of the bikeability clusters confirms results from previous research, which found that dense urban networks are often more conducive to active mobility \citep{nielsen_bikeability_2018}. Nevertheless, the bikeability clusters also show that high/low bikeability is not simply a question of urban and non-urban areas. For example, some low population areas (cluster 2) have a backbone of low-stress infrastructure and allow for low-stress cycling locally, however, usually without low-stress infrastructure connecting to locations further away. Other, also predominantly non-urban areas (cluster 3) stand out as surprisingly well-connected, despite having a low-density of low-stress infrastructure locally. Here, we often see low-stress connections that span more than \num[round-mode=places,round-precision=0]{5} km.

\section{Discussion}
\label{section:discussion}

In the following section, we discuss limitations in the method, examine implications of the results, issue recommendations for future research, and suggest policies to reduce traffic stress.

Taking the entire country of Denmark as the study area limits some of the common issues in network analysis with edge effects and sensitivity to the delimitation of the study area \citep{knight_metrics_2015, gil_street_2017}, except along natural and country borders. However, the large study area can also hide smaller local variations. For example, smaller differences in access to low-stress infrastructure can have a large impact at the local level, but seem insignificant when compared to much larger differences in low-stress infrastructure at the regional or national scale. Additionally, the clustering approach can disguise in-cluster variability. It is thus important to remember that the bikeability clusters are constructed and ranked at a country level, and that a similar analysis at e.g.~the municipal scale likely would reveal important local differences. Furthermore, in this study we simply distinguish between urban- and non-urban areas based on available data on urban zones. In practice, the distinction between urban and non-urban areas includes more complex factors. Future research could further investigate how bikeability varies across different types of urban topographies.

Although we have put a great deal of effort into refining and updating the road network data used in this study, data availability and data quality remain some of the biggest challenges for this study in particular and for research on active mobility in general \citep{willberg_comparing_2021, hill_integrated_2024}. While OSM data are generally of high quality, and for bicycle infrastructure specifically is better or comparable to official data sources \citep{hochmair_assessing_2015, ferster_using_2020, viero_how_2024}, data quality for many of the OSM tags used in the LTS classification are unknown. The use of data interpolation for missing values introduces additional uncertainties, especially in rural areas \citep{wang_investigating_2022} where the LTS classification in some locations simply becomes a proxy for road type. The received feedback on the LTS classification used in this study confirmed this tendency, with more positive feedback on accuracy in urban areas and larger disagreements on LTS classifications for municipalities in lower-density areas. Due to a lack of available data, important factors such as on-street parking, width of bicycle tracks and lanes, and traffic volumes are not included in the analysis. In particular, the lack of traffic volume data entails that some roads might receive a better or worse LTS score than warranted, if the road has more or less traffic than suggested by its road type. The issue is particularly a challenge for areas without a lot of heavy traffic, such as the many small Danish islands, which might be given a better bikeability ranking if traffic volume data were available. Similarly, we do not include data on bicycle volumes, although crowding on bicycle tracks is a concern for many cyclists \citep{uijtdewilligen_effects_2024}.

Furthermore, classifying a road as high-stress does not entail that nobody will or can cycle on that road, or that it \textit{always} will be unsafe. Rather, it means that the road characteristics are associated with cyclists feeling unsafe and stressful. Perception of safety is important: it often shapes people's choices regarding cycling, and people who report feeling unsafe are significantly less likely to bike \citep{martens_2_2019, schmidt_identifying_2024}. Although there is no doubt that e.g.~high traffic speeds not only make cyclists feel unsafe but also increase the risk of traffic deaths manifold \citep{isaksson-hellman_effect_2019}, any road design that makes cyclists feel stressed or unsafe is a barrier to more active mobility.

The study does not include data on intersections, which poses another limitation to the results. OSM in general does not have consistent and sufficiently detailed data on intersections to accurately classify them as low or high-stress, just as there are disagreements on the safety and comfort of specific intersection designs \citep{carter_bicyclist_2007, friel_cyclists_2023, werner_bikeability_2024}. The lack of intersection data is especially a concern in urban areas, where most traffic collisions involving cyclists occur at intersections \citep{isaksson-hellman_study_2012}, while bicycle crashes in rural areas often occur along the road \citep{gardner_preliminary_1998, bella_interaction_2017}.
 
The lack of high-quality data for active mobility research is not just a question of accuracy of LTS classifications. Rather, it is part of a larger structural issue, where a lack of funding and commitment to collect and improve data of relevance to active mobility undermines efforts in both research and policy making, mirroring existing inequalities in resource and space allocation for non-motorized transport \citep{behrendt_mobility_2024}. An important question for researchers and planners is thus how we can improve data on cycling and cycling conditions, particularly in non-urban areas. 

Another future research question raised by the results in this study is a further exploration of spatial equity and justice in relation to the distribution of low-stress infrastructure, such as how it relates to income, car ownership or access to alternative modes of transport. In this study, we have shown that high bikeability correlates with high population densities, and that a large share of the Danish population thus lives in high or medium bikeable areas. While e.g.~having more protected bicycle tracks in high-density areas makes sense from a utilitarian perspective, it is important to remember that this still leaves a large part of the population with very limited access to low-stress infrastructure ($\sim$\num[round-mode=places,round-precision=0]{43}\% of the population lives in cluster 1 and 2). These are predominantly rural areas, which is also where cycling rates are lowest and declining  \citep{rich_our_2023}. Rural and suburban areas also frequently have limited access to public transport. From a redistributive or minimum accessibility perspective \citep{pereira_transportation_2021}, resources and efforts to improve cycling conditions should therefore (also) be directed toward rural and suburban areas, despite their lower population densities. Additionally, it is worth noting that adding low stress bicycle infrastructure not only benefits local residents, but increases mobility for everyone cycling through the area, including bicycle tourists \citep{dansk_kyst-_og_naturturisme_metodehandbog_2024}. A detailed discussion of the equity aspects of the distribution of low and high bikeability is outside the scope of this study, but the findings nevertheless point to important barriers to increasing rural cycling.

Finally, the end goal is not just to \textit{understand} cycling conditions, but to \textit{improve} them. Therefore, future research should importantly cover how traffic stress best and most efficiently can be decreased, not least outside of high-density urban areas. Recent research has shown a wide range of positive benefits from decreasing speed limits in cities across Europe \citep{yannis_review_2024}. Although enforcing lower speed limits might be more complicated in rural areas, lowering speed limits is an effective way of reducing traffic stress and traffic collisions \citep{isaksson-hellman_effect_2019, phuksuksakul_role_2024, yannis_review_2024}.

\section{Conclusion}
\label{section:conclusion}

In this study, we have examined bikeability across Denmark with the aim of identifying spatial patterns in access to good cycling conditions. The analysis shows that bikeability -- measured as the interplay between network density, network fragmentation, and network reach for different levels of traffic stress -- varies substantially between different parts of the country with high bikeability concentrated in the most densely populated areas. The spatial concentration of low-stress infrastructure contributes to a highly fragmented network, with islands of low-stress infrastructure only connected by roads with much higher levels of traffic stress. We conclude that bikeability tends to be higher in densely populated areas, confirming findings from related research. This finding suggests the prevalence of a utilitarian prioritization of resources for bicycle infrastructure towards urban locations with the most cyclists. The results additionally indicate that many suburban and rural locations do not yet have the necessary bicycle infrastructure to support a transition toward more active mobility. To support higher cycling rates outside of urban centers, considerations of spatial equity and minimum thresholds for access to low-stress bicycle infrastructure should guide future investments and policies. Finally, the analysis also identified suburban and rural locations with high shares of low-stress infrastructure and good low-stress reach along some rural roads, where bicycle tracks have been installed. These findings highlight the effect of strategic additions of dedicated bicycle infrastructure, such as cycle highways, and illustrate that it is indeed possible to bridge urban-rural divides in bikeability.

\section*{Acknowledgments}
Data from © OpenStreetMap contributors © GeoDanmark © SDFI © Statistics Denmark © European Commission. Thanks to all OSM contributors for helping make spatial data open and free. We acknowledge support by the Danish Ministry of Transport (Grant number: CP21-033).
 
\section*{Conflicts of interest}
The authors declare no conflict of interest.

\section*{Reproducibility and data availability}

All data, code, and documentation used in the analysis and needed to reproduce the results can be found at ~\url{https://github.com/anerv/dk_bicycle_network} and \url{https://github.com/anerv/dk_network_analysis}. The data processing and analysis are customized to the Danish context and local data, but can be adapted to other locations. 

\newpage

\bibliographystyle{apalike}
\bibliography{references}

\begin{thebibliography}{}

\bibitem[Anderson, 2019]{anderson_dataanalyse_2019}
Anderson, M.~K. (2019).
\newblock Dataanalyse og modellering af transportadfærden i {Moving} {People} projektet - {Forskeranalysen}, {Moving} {People}.
\newblock Technical report, DTU, Lyngby, Denmark.

\bibitem[Anselin, 1995]{anselin_local_1995}
Anselin, L. (1995).
\newblock Local {Indicators} of {Spatial} {Association}—{LISA}.
\newblock {\em Geographical Analysis}, 27(2):93--115.
\newblock \_eprint: https://onlinelibrary.wiley.com/doi/pdf/10.1111/j.1538-4632.1995.tb00338.x.

\bibitem[Arellana et~al., 2020]{arellana_developing_2020}
Arellana, J., Saltarín, M., Larrañaga, A.~M., González, V.~I., and Henao, C.~A. (2020).
\newblock Developing an urban bikeability index for different types of cyclists as a tool to prioritise bicycle infrastructure investments.
\newblock {\em Transportation Research Part A: Policy and Practice}, 139:310--334.

\bibitem[Assunçao-Denis and Tomalty, 2019]{assuncao-denis_increasing_2019}
Assunçao-Denis, M.-E. and Tomalty, R. (2019).
\newblock Increasing cycling for transportation in {Canadian} communities: {Understanding} what works.
\newblock {\em Transportation Research Part A: Policy and Practice}, 123:288--304.

\bibitem[Aytur et~al., 2011]{aytur_pedestrian_2011}
Aytur, S.~A., Satinsky, S.~B., Evenson, K.~R., and Rodríguez, D.~A. (2011).
\newblock Pedestrian and bicycle planning in rural communities: tools for active living.
\newblock {\em Family \& Community Health}, 34(2):173--181.

\bibitem[Barrington-Leigh and Millard-Ball, 2017]{barrington-leigh_worlds_2017}
Barrington-Leigh, C. and Millard-Ball, A. (2017).
\newblock The world's user-generated road map is more than 80\% complete.
\newblock {\em PloS one}, 12(8):e0180698.
\newblock Publisher: Public Library of Science.

\bibitem[Bearn et~al., 2018]{bearn_adaption_2018}
Bearn, C., Mingus, C., and Watkins, K. (2018).
\newblock An adaption of the level of traffic stress based on evidence from the literature and widely available data.
\newblock {\em Research in Transportation Business \& Management}, 29:50--62.

\bibitem[Beecham et~al., 2023]{beecham_connected_2023}
Beecham, R., Yang, Y., Tait, C., and Lovelace, R. (2023).
\newblock Connected bike ability in {London}: {Which} localities are better connected by bike and does this matter?
\newblock {\em Environment and Planning B: Urban Analytics and City Science}, page 23998083231165122.
\newblock Publisher: SAGE Publications Ltd STM.

\bibitem[Behrendt and Sheller, 2024]{behrendt_mobility_2024}
Behrendt, F. and Sheller, M. (2024).
\newblock Mobility data justice.
\newblock {\em Mobilities}, 19(1):151--169.
\newblock Publisher: Routledge \_eprint: https://doi.org/10.1080/17450101.2023.2200148.

\bibitem[Bella and Silvestri, 2017]{bella_interaction_2017}
Bella, F. and Silvestri, M. (2017).
\newblock Interaction driver–bicyclist on rural roads: {Effects} of cross-sections and road geometric elements.
\newblock {\em Accident Analysis \& Prevention}, 102:191--201.

\bibitem[Berrigan et~al., 2010]{berrigan_associations_2010}
Berrigan, D., Pickle, L.~W., and Dill, J. (2010).
\newblock Associations between street connectivity and active transportation.
\newblock {\em International Journal of Health Geographics}, 9:20.

\bibitem[Borgatti, 2005]{borgatti_centrality_2005}
Borgatti, S.~P. (2005).
\newblock Centrality and network flow.
\newblock {\em Social Networks}, 27(1):55--71.

\bibitem[Bourne et~al., 2020]{bourne_impact_2020}
Bourne, J.~E., Cooper, A.~R., Kelly, P., Kinnear, F.~J., England, C., Leary, S., and Page, A. (2020).
\newblock The impact of e-cycling on travel behaviour: {A} scoping review.
\newblock {\em Journal of Transport \& Health}, 19:100910.

\bibitem[Buehler and Dill, 2016]{buehler_bikeway_2016}
Buehler, R. and Dill, J. (2016).
\newblock Bikeway {Networks}: {A} {Review} of {Effects} on {Cycling}.
\newblock {\em Transport Reviews}, 36(1):9--27.
\newblock Publisher: Routledge \_eprint: https://doi.org/10.1080/01441647.2015.1069908.

\bibitem[Buehler and Pucher, 2012]{buehler_cycling_2012}
Buehler, R. and Pucher, J. (2012).
\newblock Cycling to work in 90 large {American} cities: new evidence on the role of bike paths and lanes.
\newblock {\em Transportation}, 39(2):409--432.

\bibitem[Cabral and Kim, 2022]{cabral_empirical_2022}
Cabral, L. and Kim, A.~M. (2022).
\newblock An empirical reappraisal of the level of traffic stress framework for segments.
\newblock {\em Travel Behaviour and Society}, 26:143--158.

\bibitem[Carstensen and Ebert, 2012]{carstensen_cycling_2012}
Carstensen, T.~A. and Ebert, A.-K. (2012).
\newblock Cycling {Cultures} in {Northern} {Europe}: {From} ‘{Golden} {Age}’ to ‘{Renaissance}’.
\newblock In Parkin, J., editor, {\em Cycling and {Sustainability}}, volume~1 of {\em Transport and {Sustainability}}, pages 23--58. Emerald Group Publishing Limited.

\bibitem[Carter et~al., 2007]{carter_bicyclist_2007}
Carter, D.~L., Hunter, W.~W., Zegeer, C.~V., Stewart, J.~R., and Huang, H. (2007).
\newblock Bicyclist {Intersection} {Safety} {Index}.
\newblock {\em Transportation Research Record}, 2031(1):18--24.
\newblock Publisher: SAGE Publications Inc.

\bibitem[Castañon and Ribeiro, 2021]{castanon_bikeability_2021}
Castañon, U.~N. and Ribeiro, P. J.~G. (2021).
\newblock Bikeability and {Emerging} {Phenomena} in {Cycling}: {Exploratory} {Analysis} and {Review}.
\newblock {\em Sustainability}, 13(4):2394.
\newblock Number: 4 Publisher: Multidisciplinary Digital Publishing Institute.

\bibitem[Cervero et~al., 2019]{cervero_network_2019}
Cervero, R., Denman, S., and Jin, Y. (2019).
\newblock Network design, built and natural environments, and bicycle commuting: {Evidence} from {British} cities and towns.
\newblock {\em Transport Policy}, 74:153--164.

\bibitem[Chen et~al., 2017]{chen_how_2017}
Chen, C., Anderson, J.~C., Wang, H., Wang, Y., Vogt, R., and Hernandez, S. (2017).
\newblock How bicycle level of traffic stress correlate with reported cyclist accidents injury severities: {A} geospatial and mixed logit analysis.
\newblock {\em Accident; Analysis and Prevention}, 108:234--244.

\bibitem[Christiansen and Baescu, 2023]{christiansen_danish_2023}
Christiansen, H. and Baescu, O. (2023).
\newblock The {Danish} {National} {Travel} {Survey}: {Annual} {Statistical} {Report} for {Denmark} for 2022.
\newblock Technical report, DTU, Lyngby, Denmark.

\bibitem[Christiansen et~al., 2016]{christiansen_international_2016}
Christiansen, L.~B., Cerin, E., Badland, H., Kerr, J., Davey, R., Troelsen, J., van Dyck, D., Mitáš, J., Schofield, G., Sugiyama, T., Salvo, D., Sarmiento, O.~L., Reis, R., Adams, M., Frank, L., and Sallis, J.~F. (2016).
\newblock International comparisons of the associations between objective measures of the built environment and transport-related walking and cycling: {IPEN} adult study.
\newblock {\em Journal of Transport \& Health}, 3(4):467--478.

\bibitem[Clayton et~al., 2017]{clayton_cycling_2017}
Clayton, W., Parkin, J., and Billington, C. (2017).
\newblock Cycling and disability: {A} call for further research.
\newblock {\em Journal of Transport \& Health}, 6:452--462.

\bibitem[Crist et~al., 2019]{crist_fear_2019}
Crist, K., Schipperijn, J., Ryan, S., Appleyard, B., Godbole, S., and Kerr, J. (2019).
\newblock Fear {Factor}: {Level} of {Traffic} {Stress} and {GPS} {Assessed} {Cycling} {Routes}.
\newblock {\em Journal of Transportation Technologies}, 9(1):14--30.
\newblock Number: 1 Publisher: Scientific Research Publishing.

\bibitem[{Danmarks Miljøportal}, 2024]{danmarks_miljoportal_planlaegning_2024}
{Danmarks Miljøportal} (2024).
\newblock Planlægning zonekort.

\bibitem[{Dansk Kyst- og Naturturisme}, 2024]{dansk_kyst-_og_naturturisme_metodehandbog_2024}
{Dansk Kyst- og Naturturisme} (2024).
\newblock Metodehåndbog - {Kommunal} kvalificering af {Danmarks} rekreative cykelnetværk. {Fra} planlægningsnetværk til digital visning.
\newblock Technical report, Dansk Kyst- og Naturturisme, Aabybro.

\bibitem[Dill, 2004]{dill_measuring_2004}
Dill, J. (2004).
\newblock Measuring {Network} {Connectivity} for {Bicycling} and {Walking}.
\newblock In {\em 83rd {Annual} {Meeting} of the {Transportation} {Research} {Board}}, volume~83, page~19, Washington, DC.

\bibitem[Dill and Carr, 2003]{dill_bicycle_2003}
Dill, J. and Carr, T. (2003).
\newblock Bicycle {Commuting} and {Facilities} in {Major} {U}.{S}. {Cities}: {If} {You} {Build} {Them}, {Commuters} {Will} {Use} {Them}.
\newblock {\em Transportation Research Record}, 1828(1):116--123.
\newblock Publisher: SAGE Publications Inc.

\bibitem[Dill and McNeil, 2016]{dill_revisiting_2016}
Dill, J. and McNeil, N. (2016).
\newblock Revisiting the {Four} {Types} of {Cyclists}: {Findings} from a {National} {Survey}.
\newblock {\em Transportation Research Record}, 2587(1):90--99.
\newblock Publisher: SAGE Publications Inc.

\bibitem[Doran et~al., 2021]{doran_pursuit_2021}
Doran, A., El-Geneidy, A., and Manaugh, K. (2021).
\newblock The pursuit of cycling equity: {A} review of {Canadian} transport plans.
\newblock {\em Journal of Transport Geography}, 90:102927.

\bibitem[Faghih~Imani et~al., 2019]{faghih_imani_cycle_2019}
Faghih~Imani, A., Miller, E.~J., and Saxe, S. (2019).
\newblock Cycle accessibility and level of traffic stress: {A} case study of {Toronto}.
\newblock {\em Journal of Transport Geography}, 80:102496.

\bibitem[Feng and Zhang, 2019]{feng_algorithms_2019}
Feng, C. and Zhang, W. (2019).
\newblock Algorithms for the parametric analysis of metric, directional, and intersection reach.
\newblock {\em Environment and Planning B: Urban Analytics and City Science}, 46(8):1422--1438.
\newblock Publisher: SAGE Publications Ltd STM.

\bibitem[Ferster et~al., 2020]{ferster_using_2020}
Ferster, C., Fischer, J., Manaugh, K., Nelson, T., and Winters, M. (2020).
\newblock Using {OpenStreetMap} to inventory bicycle infrastructure: {A} comparison with open data from cities.
\newblock {\em International Journal of Sustainable Transportation}, 14(1):64--73.

\bibitem[Ferster et~al., 2023]{ferster_developing_2023}
Ferster, C., Nelson, T., Manaugh, K., Beairsto, J., Laberee, K., and Winters, M. (2023).
\newblock Developing a national dataset of bicycle infrastructure for {Canada} using open data sources.
\newblock {\em Environment and Planning B: Urban Analytics and City Science}, page 23998083231159905.
\newblock Publisher: SAGE Publications Ltd STM.

\bibitem[Filho et~al., 2024]{filho_cycle_2024}
Filho, F. E.~M., Ploegmakers, H., and de~Kruijf, J. (2024).
\newblock Cycle highway effects: {Assessing} modal choice to cycling in the {Netherlands}.
\newblock {\em Transportation Research Part A: Policy and Practice}, 189:104216.
\newblock Publisher: Pergamon.

\bibitem[Fonte et~al., 2017]{fonte_assessing_2017}
Fonte, C.~C., Antoniou, V., Bastin, L., Estima, J., Jokar~Arsanjani, J., Laso~Bayas, J.~C., See, L., and Vatseva, R. (2017).
\newblock Assessing {VGI} {Data} {Quality}.
\newblock In Foody, G.~M., See, L., Fritz, S., Mooney, P., Olteanu-Raimond, A.-M., Fonte, C.~C., and Antoniou, V., editors, {\em Mapping and the {Citizen} {Sensor}}, pages 137--163. Ubiquity Press, London.

\bibitem[Fosgerau et~al., 2023]{fosgerau_bikeability_2023}
Fosgerau, M., Łukawska, M., Paulsen, M., and Rasmussen, T.~K. (2023).
\newblock Bikeability and the induced demand for cycling.
\newblock {\em Proceedings of the National Academy of Sciences}, 120(16):e2220515120.
\newblock Publisher: Proceedings of the National Academy of Sciences.

\bibitem[Friel et~al., 2023]{friel_cyclists_2023}
Friel, D., Wachholz, S., Werner, T., Zimmermann, L., Schwedes, O., and Stark, R. (2023).
\newblock Cyclists’ perceived safety on intersections and roundabouts – {A} qualitative bicycle simulator study.
\newblock {\em Journal of Safety Research}, 87:143--156.

\bibitem[Furth et~al., 2016]{furth_network_2016}
Furth, P.~G., Mekuria, M.~C., and Nixon, H. (2016).
\newblock Network {Connectivity} for {Low}-{Stress} {Bicycling}.
\newblock {\em Transportation Research Record}, 2587(1):41--49.
\newblock Publisher: SAGE Publications Inc.

\bibitem[Gardner and Gray, 1998]{gardner_preliminary_1998}
Gardner and Gray (1998).
\newblock A preliminary review of rural cycling.
\newblock Technical report, Transport Research Laboratory - Department of the Environment, Transport and the Regions.

\bibitem[GeoDanmark, 2023]{geodanmark_danmarks_2023}
GeoDanmark (2023).
\newblock Danmarks {Geografi} - {GeoDanmark}.

\bibitem[Geofabrik, 2020]{geofabrik_our_2020}
Geofabrik (2020).
\newblock Our {Download} {Server}.

\bibitem[Gil, 2017]{gil_street_2017}
Gil, J. (2017).
\newblock Street network analysis “edge effects”: {Examining} the sensitivity of centrality measures to boundary conditions.
\newblock {\em Environment and Planning B: Urban Analytics and City Science}, 44(5):819--836.
\newblock Publisher: SAGE Publications Ltd STM.

\bibitem[Gössling and McRae, 2022]{gossling_subjectively_2022}
Gössling, S. and McRae, S. (2022).
\newblock Subjectively safe cycling infrastructure: {New} insights for urban designs.
\newblock {\em Journal of Transport Geography}, 101:103340.

\bibitem[Hallberg et~al., 2021]{hallberg_modelling_2021}
Hallberg, M., Rasmussen, T.~K., and Rich, J. (2021).
\newblock Modelling the impact of cycle superhighways and electric bicycles.
\newblock {\em Transportation Research Part A: Policy and Practice}, 149:397--418.

\bibitem[Hill et~al., 2024]{hill_integrated_2024}
Hill, C., Young, M., Blainey, S., Cavazzi, S., Emberson, C., and Sadler, J. (2024).
\newblock An integrated geospatial data model for active travel infrastructure.
\newblock {\em Journal of Transport Geography}, 117:103889.

\bibitem[Hochmair et~al., 2015]{hochmair_assessing_2015}
Hochmair, H.~H., Zielstra, D., and Neis, P. (2015).
\newblock Assessing the {Completeness} of {Bicycle} {Trail} and {Lane} {Features} in {OpenStreetMap} for the {United} {States}: {Completeness} of {Bicycle} {Features} in {OpenStreetMap}.
\newblock {\em Transactions in GIS}, 19(1):63--81.

\bibitem[Houde et~al., 2018]{houde_ride_2018}
Houde, M., Apparicio, P., and Séguin, A.-M. (2018).
\newblock A ride for whom: {Has} cycling network expansion reduced inequities in accessibility in {Montreal}, {Canada}?
\newblock {\em Journal of Transport Geography}, 68:9--21.

\bibitem[Isaksson-Hellman, 2012]{isaksson-hellman_study_2012}
Isaksson-Hellman, I. (2012).
\newblock A {Study} of {Bicycle} and {Passenger} {Car} {Collisions} {Based} on {Insurance} {Claims} {Data}.
\newblock {\em Annals of Advances in Automotive Medicine / Annual Scientific Conference}, 56:3--12.

\bibitem[Isaksson-Hellman and Töreki, 2019]{isaksson-hellman_effect_2019}
Isaksson-Hellman, I. and Töreki, J. (2019).
\newblock The effect of speed limit reductions in urban areas on cyclists’ injuries in collisions with cars.
\newblock {\em Traffic Injury Prevention}, 20(sup3):39--44.
\newblock Publisher: Taylor \& Francis \_eprint: https://doi.org/10.1080/15389588.2019.1680836.

\bibitem[Jones and Carlson, 2003]{jones_development_2003}
Jones, E.~G. and Carlson, T.~D. (2003).
\newblock Development of {Bicycle} {Compatibility} {Index} for {Rural} {Roads} in {Nebraska}.
\newblock {\em Transportation Research Record}, 1828(1):124--132.
\newblock Publisher: SAGE Publications Inc.

\bibitem[Jordahl et~al., 2021]{jordahl_geopandasgeopandas_2021}
Jordahl, K., Bossche, J. V.~d., Fleischmann, M., McBride, J., Wasserman, J., Badaracco, A.~G., Gerard, J., Snow, A.~D., Tratner, J., Perry, M., Farmer, C., Hjelle, G.~A., Cochran, M., Gillies, S., Culbertson, L., Bartos, M., Ward, B., Caria, G., Taves, M., Eubank, N., sangarshanan, Flavin, J., Richards, M., Rey, S., maxalbert, Bilogur, A., Ren, C., Arribas-Bel, D., Mesejo-León, D., and Wasser, L. (2021).
\newblock geopandas/geopandas: v0.10.2.

\bibitem[Kamel and Sayed, 2021]{kamel_impact_2021}
Kamel, M.~B. and Sayed, T. (2021).
\newblock The impact of bike network indicators on bike kilometers traveled and bike safety: {A} network theory approach.
\newblock {\em Environment and Planning B: Urban Analytics and City Science}, 48(7):2055--2072.
\newblock Publisher: SAGE Publications Ltd STM.

\bibitem[Kellstedt et~al., 2021]{kellstedt_scoping_2021}
Kellstedt, D.~K., Spengler, J.~O., Foster, M., Lee, C., and Maddock, J.~E. (2021).
\newblock A {Scoping} {Review} of {Bikeability} {Assessment} {Methods}.
\newblock {\em Journal of Community Health}, 46(1):211--224.

\bibitem[Kircher et~al., 2022]{kircher_cycling_2022}
Kircher, K., Forward, S., and Wallén~Warner, H. (2022).
\newblock {\em Cycling in rural areas : an overview of national and international literature}.
\newblock Statens väg- och transportforskningsinstitut.

\bibitem[Knight and Marshall, 2015]{knight_metrics_2015}
Knight, P.~L. and Marshall, W.~E. (2015).
\newblock The metrics of street network connectivity: their inconsistencies.
\newblock {\em Journal of Urbanism: International Research on Placemaking and Urban Sustainability}, 8(3):241--259.
\newblock Publisher: Routledge \_eprint: https://doi.org/10.1080/17549175.2014.909515.

\bibitem[Krizek and Roland, 2005]{krizek_what_2005}
Krizek, K.~J. and Roland, R.~W. (2005).
\newblock What is at the end of the road? {Understanding} discontinuities of on-street bicycle lanes in urban settings.
\newblock {\em Transportation Research Part D: Transport and Environment}, 10(1):55--68.

\bibitem[Lee et~al., 2017]{lee_understanding_2017}
Lee, R.~J., Sener, I.~N., and Jones, S.~N. (2017).
\newblock Understanding the role of equity in active transportation planning in the {United} {States}.
\newblock {\em Transport Reviews}, 37(2):211--226.
\newblock Publisher: Routledge \_eprint: https://doi.org/10.1080/01441647.2016.1239660.

\bibitem[Leichenko and Taylor, 2024]{leichenko_promoting_2024}
Leichenko, R. and Taylor, C. (2024).
\newblock Promoting rural sustainability transformations: {Insights} from {U}.{S}. bicycle route and trail studies.
\newblock {\em Journal of Rural Studies}, 106:103205.

\bibitem[Lowry and Loh, 2017]{lowry_quantifying_2017}
Lowry, M. and Loh, T.~H. (2017).
\newblock Quantifying bicycle network connectivity.
\newblock {\em Preventive Medicine}, 95:S134--S140.

\bibitem[Lowry et~al., 2012]{lowry_assessment_2012}
Lowry, M.~B., Callister, D., Gresham, M., and Moore, B. (2012).
\newblock Assessment of {Communitywide} {Bikeability} with {Bicycle} {Level} of {Service}.
\newblock {\em Transportation Research Record}, 2314(1):41--48.
\newblock Publisher: SAGE Publications Inc.

\bibitem[Lucas-Smith, 2019]{lucas-smith_is_2019}
Lucas-Smith, M. (2019).
\newblock Is the {OSM} data model creaking?

\bibitem[Macpherson et~al., 2004]{macpherson_urbanrural_2004}
Macpherson, A.~K., To, T.~M., Parkin, P.~C., Moldofsky, B., Wright, J.~G., Chipman, M.~L., and Macarthur, C. (2004).
\newblock Urban/rural variation in children’s bicycle-related injuries.
\newblock {\em Accident Analysis \& Prevention}, 36(4):649--654.

\bibitem[Marshall et~al., 2018]{marshall_street_2018}
Marshall, S., Gil, J., Kropf, K., Tomko, M., and Figueiredo, L. (2018).
\newblock Street network studies: from networks to models and their representations.
\newblock {\em Networks and Spatial Economics}, 18(3):735--749.
\newblock Publisher: Springer.

\bibitem[Martens et~al., 2019]{martens_2_2019}
Martens, K., Bastiaanssen, J., and Lucas, K. (2019).
\newblock 2 - {Measuring} transport equity: {Key} components, framings and metrics.
\newblock In Lucas, K., Martens, K., Di~Ciommo, F., and Dupont-Kieffer, A., editors, {\em Measuring {Transport} {Equity}}, pages 13--36. Elsevier.

\bibitem[McAndrews et~al., 2017]{mcandrews_reach_2017}
McAndrews, C., Okuyama, K., and Litt, J.~S. (2017).
\newblock The {Reach} of {Bicycling} in {Rural}, {Small}, and {Low}-{Density} {Places}.
\newblock {\em Transportation Research Record: Journal of the Transportation Research Board}, 2662(1):134--142.

\bibitem[McAndrews et~al., 2018]{mcandrews_motivations_2018}
McAndrews, C., Tabatabaie, S., and Litt, J.~S. (2018).
\newblock Motivations and {Strategies} for {Bicycle} {Planning} in {Rural}, {Suburban}, and {Low}-{Density} {Communities}: {The} {Need} for {New} {Best} {Practices}.
\newblock {\em Journal of the American Planning Association}, 84(2):99--111.
\newblock Publisher: Routledge \_eprint: https://doi.org/10.1080/01944363.2018.1438849.

\bibitem[Mekuria et~al., 2012]{mekuria_low-stress_2012}
Mekuria, M.~C., Furth, P.~G., and Nixon, H. (2012).
\newblock Low-{Stress} {Bicycling} and {Network} {Connectivity}.
\newblock Technical Report 11-19, Mineta Transportation Institute.

\bibitem[Mueller et~al., 2018]{mueller_health_2018}
Mueller, N., Rojas-Rueda, D., Salmon, M., Martinez, D., Ambros, A., Brand, C., de~Nazelle, A., Dons, E., Gaupp-Berghausen, M., Gerike, R., Götschi, T., Iacorossi, F., Int~Panis, L., Kahlmeier, S., Raser, E., and Nieuwenhuijsen, M. (2018).
\newblock Health impact assessment of cycling network expansions in {European} cities.
\newblock {\em Preventive Medicine}, 109:62--70.

\bibitem[Natera~Orozco et~al., 2020]{natera_orozco_data-driven_2020}
Natera~Orozco, L.~G., Battiston, F., Iñiguez, G., and Szell, M. (2020).
\newblock Data-driven strategies for optimal bicycle network growth.
\newblock {\em Royal Society Open Science}, 7(12):201130.

\bibitem[Nielsen and Skov-Petersen, 2018]{nielsen_bikeability_2018}
Nielsen, T. A.~S. and Skov-Petersen, H. (2018).
\newblock Bikeability – {Urban} structures supporting cycling. {Effects} of local, urban and regional scale urban form factors on cycling from home and workplace locations in {Denmark}.
\newblock {\em Journal of Transport Geography}, 69:36--44.

\bibitem[Noël et~al., 2003]{noel_compatibility_2003}
Noël, N., Leclerce, C., and Lee-Gosselin, M. (2003).
\newblock Compatibility of {Roads} for {Cyclists} in {Rural} and {Urban} {Fringe} {Areas}.
\newblock In {\em {TRB} 2003 {Annual} {Meeting}}.

\bibitem[Pedregosa et~al., 2011]{pedregosa_scikit-learn_2011}
Pedregosa, F., Varoquaux, G., Gramfort, A., Michel, V., Thirion, B., Grisel, O., Blondel, M., Prettenhofer, P., Weiss, R., Dubourg, V., Vanderplas, J., Passos, A., Cournapeau, D., Brucher, M., Perrot, M., and Duchesnay, E. (2011).
\newblock Scikit-learn: {Machine} {Learning} in {Python}.
\newblock {\em J. Mach. Learn. Res.}, 12(null):2825--2830.
\newblock Publisher: JMLR.org.

\bibitem[Peer et~al., 2023]{peer_which_2023}
Peer, S., Gangl, K., Spitzer, F., and van~der Werff, E. (2023).
\newblock Which policy measures can motivate active mobility in rural and semi-rural areas?
\newblock {\em Transportation Research Part D: Transport and Environment}, 118:103688.

\bibitem[Peponis et~al., 2008]{peponis_connectivity_2008}
Peponis, J., Bafna, S., and Zhang, Z. (2008).
\newblock The {Connectivity} of {Streets}: {Reach} and {Directional} {Distance}.
\newblock {\em Environment and Planning B: Planning and Design}, 35(5):881--901.
\newblock Publisher: SAGE Publications Ltd STM.

\bibitem[Pereira and Karner, 2021]{pereira_transportation_2021}
Pereira, R. H.~M. and Karner, A. (2021).
\newblock Transportation {Equity}.
\newblock In Vickerman, R., editor, {\em International {Encyclopedia} of {Transportation}}, pages 271--277. Elsevier, Oxford.

\bibitem[pgRouting Project, 2024]{pgrouting_project_pgrouting_2024}
pgRouting Project (2024).
\newblock {pgRouting} {Project} — {Open} {Source} {Routing} {Library}.

\bibitem[Phuksuksakul et~al., 2024]{phuksuksakul_role_2024}
Phuksuksakul, N., Haque, M., and Yasmin, S. (2024).
\newblock The role of posted speed limit on pedestrian and bicycle injury severities: {An} investigation into systematic and unobserved heterogeneities.
\newblock {\em Analytic Methods in Accident Research}, 44:100351.

\bibitem[PSC, 2023]{postgis_psc_postgis_2023}
PSC, P. (2023).
\newblock {PostGIS}.

\bibitem[Reggiani et~al., 2022]{reggiani_understanding_2022}
Reggiani, G., Van~Oijen, T., Hamedmoghadam, H., Daamen, W., Vu, H.~L., and Hoogendoorn, S. (2022).
\newblock Understanding bikeability: a methodology to assess urban networks.
\newblock {\em Transportation}, 49(3):897--925.

\bibitem[Reggiani et~al., 2023]{reggiani_multi-city_2023}
Reggiani, G., Verma, T., Daamen, W., and Hoogendoorn, S. (2023).
\newblock A multi-city study on structural characteristics of bicycle networks.
\newblock {\em Environment and Planning B: Urban Analytics and City Science}, 50(8):2017--2037.

\bibitem[Rey and Anselin, 2007]{rey_pysal_2007}
Rey, S.~J. and Anselin, L. (2007).
\newblock {PySAL}: {A} {Python} {Library} of {Spatial} {Analytical} {Methods}.
\newblock {\em Review of Regional Studies}, 37(1).

\bibitem[Rey et~al., 2020]{rey_geographic_2020}
Rey, S.~J., Arribas-Bel, D., and Wolf, L.~J. (2020).
\newblock Geographic {Thinking} for {Data} {Scientists} — {Geographic} {Data} {Science} with {Python}.

\bibitem[Rich et~al., 2023]{rich_our_2023}
Rich, J., Myhrmann, M.~S., and Mabit, S.~E. (2023).
\newblock Our children cycle less - {A} {Danish} pseudo-panel analysis.
\newblock {\em Journal of Transport Geography}, 106:103519.

\bibitem[Scappini et~al., 2022]{scappini_regional_2022}
Scappini, B., Zucca, V., Meloni, I., and Piras, F. (2022).
\newblock The regional cycle network of {Sardinia}: upgrading the accessibility of rural areas through a comprehensive island-wide cycle network.
\newblock {\em European Transport Research Review}, 14(1):10.

\bibitem[Schiavina et~al., 2023]{schiavina_ghs-pop_2023}
Schiavina, M., Freire, S., and MacManus, K. (2023).
\newblock {GHS}-{POP} {R2023A} - {GHS} population grid multitemporal (1975-2030).

\bibitem[Schmidt et~al., 2024]{schmidt_identifying_2024}
Schmidt, T., Top Klein-Wengel, T., Christiansen, L.~B., Elmose-Østerlund, K., and Schipperijn, J. (2024).
\newblock Identifying the potential for increasing cycling in {Denmark}: {Factors} associated with short-distance and long-distance commuter cycling.
\newblock {\em Journal of Transport \& Health}, 38:101870.

\bibitem[Schoner and Levinson, 2014]{schoner_missing_2014}
Schoner, J.~E. and Levinson, D.~M. (2014).
\newblock The missing link: bicycle infrastructure networks and ridership in 74 {US} cities.
\newblock {\em Transportation}, 41(6):1187--1204.

\bibitem[Schön et~al., 2024a]{schon_scoping_2024}
Schön, P., Heinen, E., and Manum, B. (2024a).
\newblock A scoping review on cycling network connectivity and its effects on cycling.
\newblock {\em Transport Reviews}, 0(0):1--25.
\newblock Publisher: Routledge \_eprint: https://doi.org/10.1080/01441647.2024.2337880.

\bibitem[Schön et~al., 2024b]{schon_impact_2024}
Schön, P., Heinen, E., Rangul, V., Sund, E.~R., and Manum, B. (2024b).
\newblock The impact of street network connectivity on active school travel: {Norway}’s {HUNT} study.
\newblock {\em Environment and Planning B: Urban Analytics and City Science}, page 23998083241235978.
\newblock Publisher: SAGE Publications Ltd STM.

\bibitem[Semler et~al., 2018]{semler_keys_2018}
Semler, C., Sanders, M., Buck, D., Dock, S., Cesme, B., and Wang, S. (2018).
\newblock The {Keys} to {Connectivity}: {The} {District} of {Columbia}’s {Innovative} {Approach} to {Unlocking} {Low}-{Stress} {Bicycle} {Networks}.
\newblock {\em Transportation Research Record}, 2672(36):63--72.
\newblock Publisher: SAGE Publications Inc.

\bibitem[Semler et~al., 2017]{semler_low-stress_2017}
Semler, C., Sanders, M., Buck, D., Graham, J., Pochowski, A., and Dock, S. (2017).
\newblock Low-{Stress} {Bicycle} {Network} {Mapping}: {The} {District} of {Columbia}’s {Innovative} {Approach} to {Applying} {Level} of {Traffic} {Stress}.
\newblock {\em Transportation Research Record}, 2662(1):31--40.
\newblock Publisher: SAGE Publications Inc.

\bibitem[Shi et~al., 2021]{shi_quantitative_2021}
Shi, C., Wei, B., Wei, S., Wang, W., Liu, H., and Liu, J. (2021).
\newblock A quantitative discriminant method of elbow point for the optimal number of clusters in clustering algorithm.
\newblock {\em EURASIP Journal on Wireless Communications and Networking}, 2021(1):31.

\bibitem[Skov-Petersen et~al., 2017]{skov-petersen_effects_2017}
Skov-Petersen, H., Jacobsen, J.~B., Vedel, S.~E., Thomas~Alexander, S.~N., and Rask, S. (2017).
\newblock Effects of upgrading to cycle highways - {An} analysis of demand induction, use patterns and satisfaction before and after.
\newblock {\em Journal of Transport Geography}, 64:203--210.

\bibitem[Sorton and Walsh, 1994]{sorton_bicycle_1994}
Sorton, A. and Walsh, T. (1994).
\newblock Bicycle {Stress} {Level} as a {Tool} {To} {Evaluate} {Urban} and {Suburban} {Bicycle} {Compatibility}.
\newblock {\em Transportation Research Record}, 1438.

\bibitem[{Statistics Denmark}, 2023]{statistics_denmark_mere_2023}
{Statistics Denmark} (2023).
\newblock Mere end seks ud af ti familier råder over bil.

\bibitem[Szell et~al., 2022]{szell_growing_2022}
Szell, M., Mimar, S., Perlman, T., Ghoshal, G., and Sinatra, R. (2022).
\newblock Growing urban bicycle networks.
\newblock {\em Scientific Reports}, 12(1):6765.
\newblock Number: 1 Publisher: Nature Publishing Group.

\bibitem[{The Danish Agency for Climate Data}, 2024]{sdfi_danske_2024}
{The Danish Agency for Climate Data} (2024).
\newblock Danske stednavne - {Grunddata}.

\bibitem[Tribby and Tharp, 2019]{tribby_examining_2019}
Tribby, C.~P. and Tharp, D.~S. (2019).
\newblock Examining urban and rural bicycling in the {United} {States}: {Early} findings from the 2017 {National} {Household} {Travel} {Survey}.
\newblock {\em Journal of Transport \& Health}, 13:143--149.

\bibitem[Uber, 2023]{uber_h3-py_2023}
Uber (2023).
\newblock h3-py: {Uber}'s {H3} {Hexagonal} {Hierarchical} {Geospatial} {Indexing} {System} in {Python}.
\newblock original-date: 2018-06-12T22:39:59Z.

\bibitem[Uijtdewilligen et~al., 2024]{uijtdewilligen_effects_2024}
Uijtdewilligen, T., Baran~Ulak, M., Jan~Wijlhuizen, G., and Geurs, K.~T. (2024).
\newblock Effects of crowding on route preferences and perceived safety of urban cyclists in the {Netherlands}.
\newblock {\em Transportation Research Part A: Policy and Practice}, 183:104030.

\bibitem[van~der Meer et~al., 2024]{van_der_meer_assessment_2024}
van~der Meer, L., Werner, C., and Loidl, M. (2024).
\newblock Assessment of bicycle accessibility to mobility hubs under different criteria for cycling network quality.
\newblock {\em AGILE: GIScience Series}, 5:1--7.
\newblock Publisher: Copernicus GmbH.

\bibitem[Vassi and Vlastos, 2014]{vassi_review_2014}
Vassi, A. and Vlastos, T. (2014).
\newblock A review and critical assessment of cycling infrastructures across {Europe}.
\newblock In {\em {WIT} {Transactions} on {Ecology} and the {Environment}}, pages 757--768, Siena, Italy.

\bibitem[Vedel et~al., 2017]{vedel_bicyclists_2017}
Vedel, S.~E., Jacobsen, J.~B., and Skov-Petersen, H. (2017).
\newblock Bicyclists’ preferences for route characteristics and crowding in {Copenhagen} – {A} choice experiment study of commuters.
\newblock {\em Transportation Research Part A: Policy and Practice}, 100:53--64.

\bibitem[Vierø, 2024]{viero_bicycle_2024}
Vierø, A.~R. (2024).
\newblock Bicycle {Suitability}.

\bibitem[Vierø et~al., 2024a]{viero_bikedna_2024}
Vierø, A.~R., Vybornova, A., and Szell, M. (2024a).
\newblock {BikeDNA}: {A} tool for bicycle infrastructure data and network assessment.
\newblock {\em Environment and Planning B: Urban Analytics and City Science}, 51(2):512--528.

\bibitem[Vierø et~al., 2024b]{viero_how_2024}
Vierø, A.~R., Vybornova, A., and Szell, M. (2024b).
\newblock How {Good} {Is} {Open} {Bicycle} {Network} {Data}? {A} {Countrywide} {Case} {Study} of {Denmark}.
\newblock {\em Geographical Analysis}, page gean.12400.

\bibitem[Vybornova et~al., 2022]{vybornova_automated_2022}
Vybornova, A., Cunha, T., Gühnemann, A., and Szell, M. (2022).
\newblock Automated {Detection} of {Missing} {Links} in {Bicycle} {Networks}.
\newblock {\em Geographical Analysis}, n/a(n/a).

\bibitem[Wang et~al., 2016]{wang_does_2016}
Wang, H., Palm, M., Chen, C., Vogt, R., and Wang, Y. (2016).
\newblock Does bicycle network level of traffic stress ({LTS}) explain bicycle travel behavior? {Mixed} results from an {Oregon} case study.
\newblock {\em Journal of Transport Geography}, 57:8--18.

\bibitem[Wang et~al., 2022]{wang_investigating_2022}
Wang, K., Akar, G., Cheng, L., Lee, K., and Sanders, M. (2022).
\newblock Investigating tools for evaluating service and improvement opportunities on bicycle routes in {Ohio}, {United} {States}.
\newblock {\em Multimodal Transportation}, 1(4):100040.

\bibitem[Wasserman et~al., 2019]{wasserman_evaluating_2019}
Wasserman, D., Rixey, A., Zhou, X.~E., Levitt, D., and Benjamin, M. (2019).
\newblock Evaluating {OpenStreetMap}’s {Performance} {Potential} for {Level} of {Traffic} {Stress} {Analysis}.
\newblock {\em Transportation Research Record}, 2673(4):284--294.
\newblock Publisher: SAGE Publications Inc.

\bibitem[Werner et~al., 2024]{werner_bikeability_2024}
Werner, C., van~der Meer, L., Kaziyeva, D., Stutz, P., Wendel, R., and Loidl, M. (2024).
\newblock Bikeability of road segments: {An} open, adjustable and extendible model.
\newblock {\em Journal of Cycling and Micromobility Research}, 2:100040.

\bibitem[Willberg et~al., 2021]{willberg_comparing_2021}
Willberg, E., Tenkanen, H., Poom, A., Salonen, M., and Toivonen, T. (2021).
\newblock Comparing spatial data sources for cycling studies: a review.
\newblock In {\em Transport in {Human} {Scale} {Cities}}, pages 169--187. Edward Elgar Publishing.
\newblock Section: Transport in Human Scale Cities.

\bibitem[Winters et~al., 2013]{winters_mapping_2013}
Winters, M., Brauer, M., Setton, E.~M., and Teschke, K. (2013).
\newblock Mapping {Bikeability}: {A} {Spatial} {Tool} to {Support} {Sustainable} {Travel}.
\newblock {\em Environment and Planning B: Planning and Design}, 40(5):865--883.
\newblock Publisher: SAGE Publications Ltd STM.

\bibitem[Winters et~al., 2011]{winters_motivators_2011}
Winters, M., Davidson, G., Kao, D., and Teschke, K. (2011).
\newblock Motivators and deterrents of bicycling: comparing influences on decisions to ride.
\newblock {\em Transportation}, 38(1):153--168.

\bibitem[Winters et~al., 2020]{winters_at--glance_2020}
Winters, M., Zanotto, M., and Butler, G. (2020).
\newblock At-a-glance - {The} {Canadian} {Bikeway} {Comfort} and {Safety} ({Can}-{BICS}) {Classification} {System}: a common naming convention for cycling infrastructure.
\newblock {\em Health Promotion and Chronic Disease Prevention in Canada : Research, Policy and Practice}, 40(9):288--293.

\bibitem[Winters et~al., 2022]{winters_canadian_2022}
Winters, M., Zanotto, M., and Butler, G. (2022).
\newblock The {Canadian} {Bikeway} {Comfort} and {Safety} metrics ({Can}-{BICS}): {National} measures of the bicycling environment for use in research and policy.
\newblock {\em Health Reports}, 33(82):13.

\bibitem[Wysling and Purves, 2022]{wysling_where_2022}
Wysling, L. and Purves, R.~S. (2022).
\newblock Where to improve cycling infrastructure? {Assessing} bicycle suitability and bikeability with open data in the city of {Paris}.
\newblock {\em Transportation Research Interdisciplinary Perspectives}, 15:100648.

\bibitem[Xiao et~al., 2022]{xiao_shifting_2022}
Xiao, C., Sluijs, E.~v., Ogilvie, D., Patterson, R., and Panter, J. (2022).
\newblock Shifting towards healthier transport: carrots or sticks? {Systematic} review and meta-analysis of population-level interventions.
\newblock {\em The Lancet Planetary Health}, 6(11):e858--e869.

\bibitem[Yannis and Michelaraki, 2024]{yannis_review_2024}
Yannis, G. and Michelaraki, E. (2024).
\newblock Review of {City}-{Wide} 30 km/h {Speed} {Limit} {Benefits} in {Europe}.
\newblock {\em Sustainability}, 16(11):4382.
\newblock Number: 11 Publisher: Multidisciplinary Digital Publishing Institute.

\bibitem[Łukawska et~al., 2023]{lukawska_joint_2023}
Łukawska, M., Paulsen, M., Rasmussen, T.~K., Jensen, A.~F., and Nielsen, O.~A. (2023).
\newblock A joint bicycle route choice model for various cycling frequencies and trip distances based on a large crowdsourced {GPS} dataset.
\newblock {\em Transportation Research Part A: Policy and Practice}, 176:103834.

\end{thebibliography}

\end{document}